\documentclass{paper} 

\usepackage{fullpage}
\usepackage{times}
\usepackage{amsmath}
\usepackage{amssymb}
\usepackage[ruled,vlined]{algorithm2e}
\usepackage{graphicx}
\usepackage{comment}
\usepackage{url}
\usepackage{setspace}
\usepackage{listings}
\usepackage{rotating}
\usepackage{multirow}
\usepackage{color}
\usepackage{balance}
\usepackage{pslatex}
\usepackage{cite}
\usepackage{wrapfig}
\usepackage{subfig}

\newenvironment{definition}[1][Definition]{\begin{trivlist}
\item[\hskip \labelsep {\bfseries #1}]}{\end{trivlist}}

\renewcommand{\vspace}[1]{}

\usepackage[ruled]{algorithm2e}
\renewcommand{\algorithmcfname}{ALGORITHM}
\SetAlFnt{\small}
\SetAlCapFnt{\small}
\SetAlCapNameFnt{\small}
\SetAlCapHSkip{0pt}
\IncMargin{-\parindent}

\begin{document}

\title{Beyond Reuse Distance Analysis:\\ Dynamic Analysis for Characterization of Data Locality Potential\footnote{This work is supported in part by the U.S. National Science Foundation
through awards 0811457, 0904549, 0926127, 0926687, 0926688 and 1321147, by the
U.S. Department of Energy through award DE-SC0008844, and by the
U.S. Army through contract W911NF-10-1-000.}
}

\author{
Naznin Fauzia$^1$,
Venmugil Elango$^1$,
Mahesh Ravishankar$^1$,
J.~Ramanujam$^2$,\\
Fabrice Rastello$^3$,
Atanas Rountev$^1$,
Louis-No{\"e}l Pouchet$^4$,
P.~Sadayappan$^1$\\
\small 1: The Ohio State University\hspace{1em}
2: Louisiana State University\hspace{1em}
3: INRIA / LIP, ENS Lyon\hspace{1em}
4: University of California Los Angeles\\
}

\maketitle

\begin{abstract}
Emerging computer architectures will feature drastically decreased
flops/byte (ratio of peak processing rate to memory bandwidth) as
highlighted by recent studies on Exascale architectural
trends. Further, flops are getting cheaper while the energy cost of
data movement is increasingly dominant.  The understanding and
characterization of data locality properties of computations is
critical in order to guide efforts to enhance data locality.

Reuse distance analysis of memory address traces is a valuable tool to
perform data locality characterization of programs. A single reuse
distance analysis can be used to estimate the number of cache misses
in a fully associative LRU cache of any size, thereby providing
estimates on the minimum bandwidth requirements at different levels of
the memory hierarchy to avoid being bandwidth bound.  However, such an
analysis only holds for the particular execution order that produced
the trace. It cannot estimate potential improvement in data locality
through dependence preserving transformations that change the
execution schedule of the operations in the computation.

In this article, we develop a novel dynamic analysis approach to 
characterize the inherent locality properties of a computation and
thereby assess the potential for data locality enhancement via
dependence preserving transformations.
The execution trace of a code is analyzed to extract a computational
directed acyclic graph (CDAG) of the data dependences. The CDAG is
then partitioned into convex subsets, and the convex partitioning is
used to reorder the operations in the execution trace to enhance data
locality. The approach enables us to go beyond reuse distance analysis
of a single specific order of execution of the operations of a
computation in characterization of its data locality properties.  It
can serve a valuable role in identifying promising code regions for
manual transformation, as well as assessing the effectiveness of
compiler transformations for data locality enhancement. We demonstrate
the effectiveness of the approach using a number of benchmarks,
including case studies where the potential shown by the analysis is
exploited to achieve lower data movement costs and better performance.

\end{abstract}

\section{Introduction}
\label{sec:intro}

Advances in technology over the last few decades have yielded
significantly different rates of improvement in the computational
performance of processors relative to the speed of memory access. The
Intel 80286 processor introduced in 1982 had an operation execution
latency of 320 ns and a main memory access time of 225 ns
\cite{Hennessey-Patterson2011}. The recent Intel Core i7 processor has
an operation latency of 4ns and a memory latency of 37 ns, illustrating
an order of magnitude shift in the ratio of operation latency to
memory access latency. Since processors use parallelism and pipelining
in execution of operations and for memory access, it is instructive to
also examine the trends in the peak execution throughput and memory
bandwidth for these two processors: 2 MIPS and 13 MBytes/sec for the
80286 versus 50,000 MIPS and 16,000 MBytes/sec for the Core i7. The
ratio of peak computational rate to peak memory access
bandwidth has also changed by more than an order of magnitude.

Because of the significant mismatch between computational latency and
throughput when compared to main memory latency and bandwidth, the use
of hierarchical memory systems and the exploitation of significant
data reuse in the higher (i.e., faster) levels of the memory hierarchy
is critical for high performance. Techniques such as pre-fetching and
overlap of computation with communication can be used to mitigate the
impact of high memory access latency on performance, but the mismatch
between maximum computational rate and peak memory bandwidth is much
more fundamental; \emph{the only solution is to limit the volume of data
movement to/from memory by enhancing data reuse in registers and
higher levels of the cache.}

A significant number of research efforts have focused on improving
data locality, by developing new algorithms such as the so called {\em
  communication avoiding} algorithms
%
%
%
\cite{demmel.unpub.08,BDHS11a,BDHS11b}
as well as automated compiler transformation techniques
\cite{irigoin88popl,wolf.91.pldi,kennedy.93.lcpc,uday08pldi}. With
future systems, the cost of data movement through the memory hierarchy
is expected to become even more dominant relative to the cost of
performing arithmetic operations
\cite{bergman2008exascale,fuller2011,shalf2011exascale}, both in terms
of throughput and energy. 
Optimizing data access costs will
become ever more critical in the coming years. Given the crucial
importance of optimizing data access costs in systems with
hierarchical memory, it is of great interest to develop tools and
techniques to assess the inherent data locality characteristics of
different parts of a computation, and the potential for data locality
enhancement via dependence preserving transformations.

Reuse distance (also called LRU stack distance) is a widely used
approach to model data
locality~\cite{mattson1970valuation,ding2003predicting} in
computations.  Since its introduction in 1970 by Mattson et
al.~\cite{mattson1970valuation}, reuse distance analysis has found
numerous applications in performance analysis and optimization, such
as cache miss rate
prediction~\cite{Zhong:pact03,Marin:sigmetrics2004,JiangZTS10},
program phase detection~\cite{Shen:asplos2004}, data layout
optimization~\cite{Zhong:pldi2004}, virtual memory
management~\cite{Cascaval:pact2005} and I/O performance
optimization~\cite{Jiang:TC2005}.  Defined as the number of distinct
memory references between two successive references to the same
location, reuse distance provides a quantification of the locality
present in a data reference trace.  A key property of the reuse
distance characterization of an address trace is that, for a fully
associative cache of size $S$, every reference with reuse distance $d
\leq S$ would be a hit and all others misses.  Thus, a single reuse
distance analysis of an address trace allows to estimate the
total number of hits/misses for an idealized cache of any size, from a
cumulative reuse distance histogram. In contrast, cache simulation to
determine the number of hits/misses would have to be repeated for each
cache size of interest.  Although real caches have non-unit line size
and finite associativity, the data transfer volume estimated from the
cache miss count via reuse distance analysis can serve as a
valuable estimate for any real cache.


Although reuse distance analysis has found many uses in characterizing
data locality in computations, it has a fundamental constraint: {\em
  The analysis is based on the memory address trace corresponding to
  \textbf{a particular execution order} of the operations constituting
  the computation.}  Thus, it does not in any way account for the
possibility of alternate valid execution orders for the computation
that may exploit much better data locality. While reuse distance
analysis provides a useful characterization of data locality for a
given execution trace, it fails to provide any information on the
potential for improvement in data locality that may be feasible
through valid reordering of the operations in the execution trace.
In particular, given only the reuse distance profile for the address
trace generated by execution of some code, it is not possible to
determine whether the observed locality characteristics reveal
fundamental inherent limitations of an algorithm, or are merely the
consequence of a sub-optimal implementation choice.

In this paper, we develop a novel dynamic analysis approach to provide
insights beyond that possible from standard reuse distance analysis.
The analysis seeks to characterize the \emph{inherent data locality
  potential} of the implemented algorithm, instead of the reuse
distance profile of the address trace from a specific execution order
of the constituent operations.  We develop graph partitioning
techniques that could be seen as a generalization of loop tiling, but
considering arbitrary shapes for the tiles that enable atomic execution
of tiles.
%
%
Instead of simply performing reuse distance analysis on the execution
trace of a given sequential program, we first explicitly construct a
computational directed acyclic graph (CDAG) to capture the
statement instances and their inter-dependences, then perform convex
partitioning of the CDAG to generate a modified dependence-preserving
execution order with better expected data reuse, and finally perform
reuse distance analysis for the address trace corresponding to the
modified execution order. We apply the proposed approach on a number
of benchmarks and demonstrate that it can be very effective.
This article makes the following contributions.
\begin{itemize}
\item It is the first work, to the best of our knowledge, to develop a
  dynamic analysis approach that seeks to characterize the inherent
  data locality characteristics of algorithms.
\item It develops effective algorithms to perform convex partitioning
  of CDAGs to enhance data locality. While convex partitioning of DAGs
  has previously been used for estimating parallel speedup, to our
  knowledge this is the first effort to use it for characterizing data
  locality potential.
\item It demonstrates the potential of the approach to identify
  opportunities for enhancement of data locality in existing
  implementations of computations. Thus, an analysis tool based on
  this approach to data locality characterization can be valuable to:
  (i) application developers, for comparing alternate algorithms and
  identifying parts of existing code that may have significant
  potential for data locality enhancement, and (ii) compiler writers,
  for assessing the effectiveness of a compiler's program optimization
  module in enhancing data locality.
\item It demonstrates, through case studies, the use of the new
  dynamic analysis approach in identifying opportunities for data
  locality optimization that are beyond the scope of the current
  state-of-the-art optimizing compilers.
For example, the insights
  from the analysis have resulted in the development of a 3D tiled
  version of the Floyd-Warshall all-pairs shortest paths algorithm,
  which was previously believed to be un-tileable without semantic
  information of the base algorithm.

\end{itemize}


The rest of the paper is organized as follows. Section~\ref{sec:bg}
presents background on reuse distance analysis, and a high-level
overview of the proposed approach for locality characterization.
%
%
The algorithmic details of the approach to convex partitioning of
CDAGs are provided in Section~\ref{sec:cp}.
Section~\ref{sec:experiments} presents experimental results.  Related
work is discussed in Section~\ref{sec:related}, followed by concluding
remarks in Sections~\ref{sec:discussion} and~\ref{sec:conc}.

\section{Background \& Overview of Approach}
\label{sec:bg}

\subsection{Reuse Distance Analysis}

Reuse distance analysis is a widely used metric that models data
locality~\cite{mattson1970valuation,ding2003predicting}.  The reuse
distance of a reference in a memory address trace is defined as the
number of distinct memory references between two successive references
to the same location.

\vspace{-.2cm}
\begin{figure}[h!tb]
  \begin{flushleft}
\centering
  \begin{tabular}{|c|c|c|c|c|c|c|c|c|c|c|}
  \hline
  Time & 0 & 1 & 2 & 3 & 4 & 5 & 6 & 7 & 8 & 9   \\
  \hline
  Data Ref. & d & a & c & b & c & c & e & b & a & d \\
  \hline
  Reuse Dist. & $\infty$ & $\infty$ & $\infty$ & $\infty$ & 1 & 0 & $\infty$ & 2 & 3 & 4 \\
  \hline
  \end{tabular}
  \end{flushleft}
\caption{Example data reference trace}
\label{fig:example-trace}
\end{figure}
\vspace{-.4cm}
An example data reference trace of length $N = 10$ is shown in
Fig.~\ref{fig:example-trace}.  It contains references to $M = 5$
distinct data addresses $\{ a,b,c,d,e \}$. As shown in the figure, each
reference to an address in the trace is associated with a
reuse distance.  The first time an address is referenced, its reuse
distance is $\infty$.  For all later references to an address, the
reuse distance is the number of distinct intervening addresses
referenced.  In the figure, address $c$ is referenced three
times. Since $b$ is referenced in between the first and second
references to $c$, the latter has a reuse distance of $1$. Since the
second and third references to $c$ are consecutive, without any other
distinct intervening references to any other addresses, the last
access to $c$ has a reuse distance of $0$.

A significant advantage of reuse distance analysis (RDA) is that a
single analysis of the address trace of a code's execution can be used
to estimate data locality characteristics as a function of the cache
size.  Given an execution trace, a reuse distance histogram for that
sequence is obtained as follows.  For each memory reference, $M$ in
the trace, its reuse distance is the number of distinct addresses in
the trace after the most recent access to $M$ (the reuse distance is
considered to be infinity if there was no previous reference in the
trace to $M$). The number of references in the trace with reuse
distance of 0, 1, 2, ..., are counted to form the reuse distance
histogram. A cumulative reuse distance histogram plots, as a function
of $k$, the number of references in the trace that have reuse distance
less than or equal to $k$.  The cumulative reuse distance histogram
directly provides the number of cache hits for a fully associative
cache of capacity $C$ with a LRU (Least Recently Used) replacement
policy, since the data accessed by any reference with reuse distance
less than or equal to $C$ would result in a cache hit.


\paragraph{Example} 
We use a simple example to illustrate both the benefits as well
as a significant limitation of standard RDA.
Fig.~\ref{fig:seidel1code} shows code for a simple Seidel-like spatial
sweep, with a default implementation and a fully equivalent
tiled variant, where the computation is executed in blocks.

\vspace{-.3cm}
\begin{figure}[h!tb]
\begin{minipage}[b]{.99\textwidth}
\centering
\begin{minipage}[b]{0.45\textwidth}
\begin{lstlisting}[frame=single,basicstyle=\scriptsize]
for(i = 1; i < N-1; i++)
  for(j = 1; j < N-1; j++)
    A[i][j] = A[i-1][j] + A[i][j-1];
\end{lstlisting}
\begin{center}
\end{center}
\end{minipage}
~~
\begin{minipage}[b]{0.45\textwidth}
\begin{lstlisting}[frame=single,basicstyle=\scriptsize]
/* B -> Tile size */
for(it = 1; it < N-1; it += B)
  for(jt = 1; jt < N-1; jt += B)
    for(i = it; i < min(it+B, N-1); i++)
      for(j = jt; j < min(jt+B, N-1); j++)
        A[i][j] = A[i-1][j] + A[i][j-1];
\end{lstlisting}
\begin{center}
\end{center}
\end{minipage}
\end{minipage}
\vspace{-.5cm}
\caption{\label{fig:seidel1code}Example: Single-sweep two-point Gauss-Seidel code, (a) Untiled and (b) Tiled}
\vspace{-.4cm}
\end{figure}

\begin{figure}[h!tb]
\centering\begin{minipage}[b]{0.99\textwidth}
\begin{minipage}[b]{0.32\textwidth}
\includegraphics[width=4.8cm]{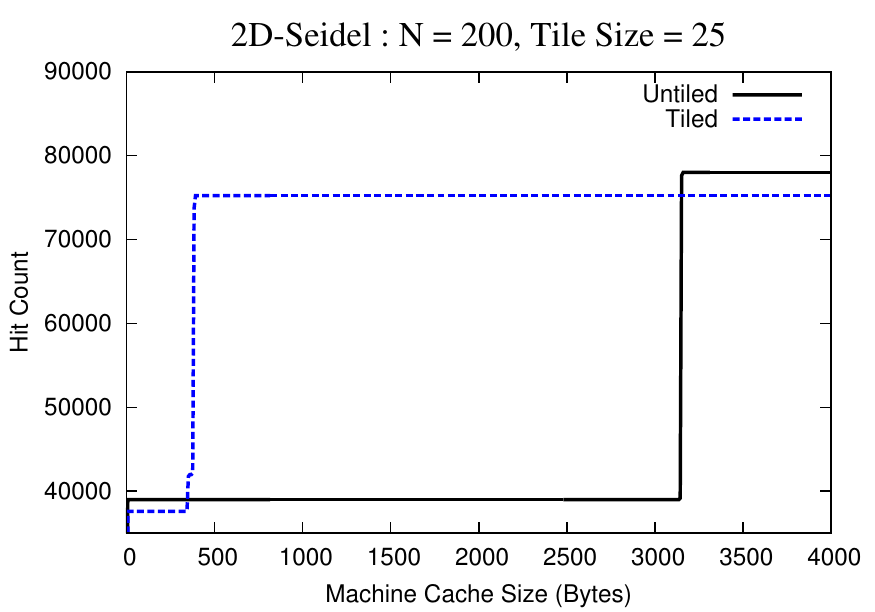}
\end{minipage}
~~
\begin{minipage}[b]{0.32\textwidth}
\includegraphics[width=4.8cm]{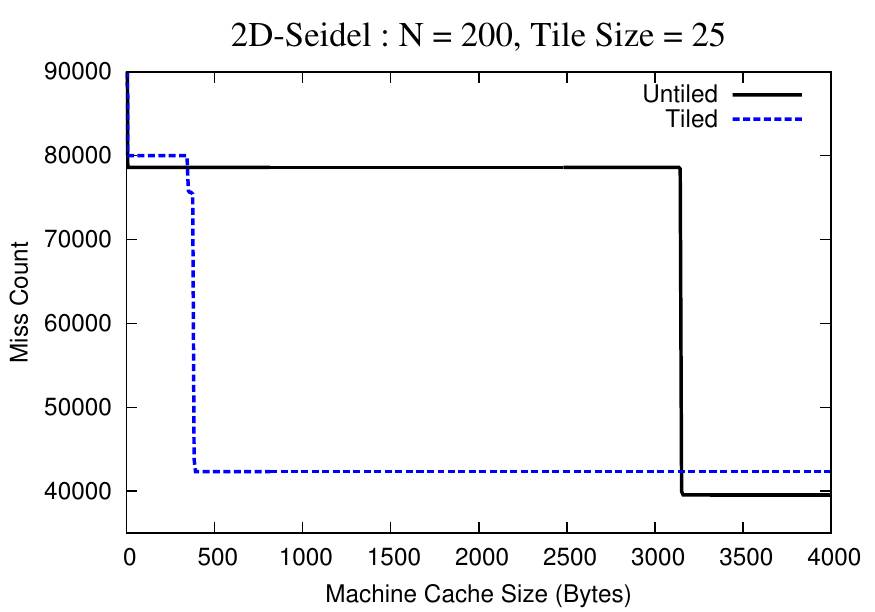}
\end{minipage}
~~
\begin{minipage}[b]{0.32\textwidth}
\includegraphics[width=4.8cm]{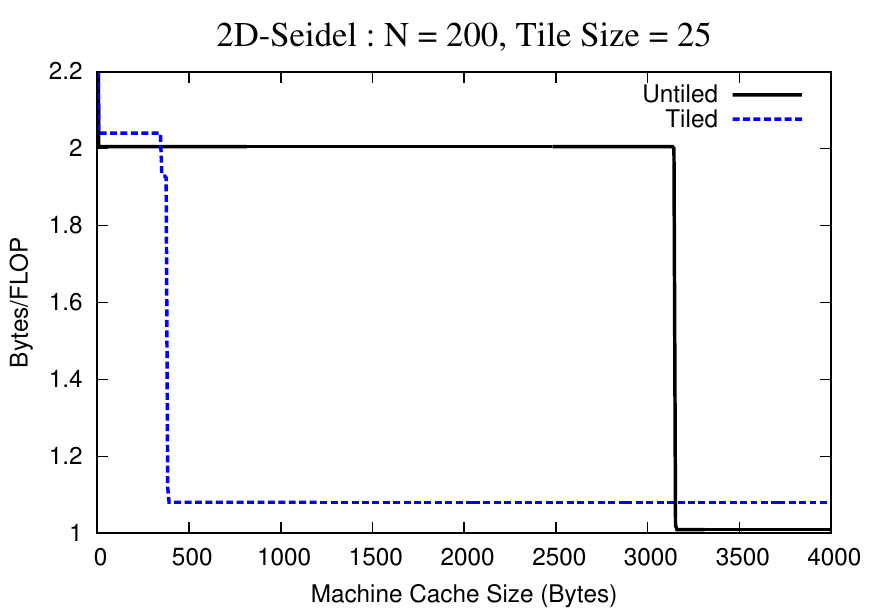}
\end{minipage}
\end{minipage}
  \caption{\label{fig:seidel1RDA} Reuse distance profile: (a) cache hit rate, (b) cache miss rate, and (c) memory bandwidth demand for tiled/untiled versions of code in Fig.~\ref{fig:seidel1code}}
\end{figure}

Fig.~\ref{fig:seidel1RDA}(a) displays the cumulative reuse distance histogram
for both versions. As explained above, it can be interpreted as
the number of data cache
hits (y axis) as a function of the cache size (x axis).
The same data is depicted in Fig.~\ref{fig:seidel1RDA}(b),
showing the number of cache misses
(by subtracting the number of hits from the total number of references).
%
%
%
%
%
untiled form of the code, a cache with capacity less than 400 words
(3200 bytes, with 8 bytes per element) will be too small to
effectively exploit reuse. The reuse distance profile for the tiled
code is quite different, suggesting that effective exploitation of
reuse is feasible with a smaller cache of capacity of 50 words (400
bytes).  This example illustrates the benefits of RDA: i) For a given
code, it provides insights into the impact of cache capacity on the
expected effectiveness of data locality exploitation, and ii) Given
two known alternative implementations for a computation, it enables a
comparative assessment of the codes with respect to data locality.

\vspace{-.4cm}
\vspace{-.7cm}

\paragraph{Limitations of reuse distance analysis} The Seidel example also 
illustrates a fundamental shortcoming of RDA that we address through
the work presented in this article: Given an execution trace for a
code, RDA only provides a data locality characterization for one {\em
  particular execution order} of the constituent operations, and
provides no insights on whether significant improvements may be
possible via dependence preserving reordering of execution of the
operations. The tiled and untiled variants in
Fig.~\ref{fig:seidel1code} represent equivalent computations, with the
only difference being the relative order of execution of exactly the
same set of primitive arithmetic operations on exactly the same sets
of array operands. Although state-of-the-art static compiler
transformation techniques (e.g., using polyhedral loop
transformations) can transform the untiled code in
Fig.~\ref{fig:seidel1code}(a) to a tiled form of
Fig.~\ref{fig:seidel1code}(b), many codes exist (as illustrated
through case studies later in this article), where data locality
characteristics can be improved, but are beyond the scope of the most
advanced compilers today. The main question that RDA does not answer
is whether the poor reuse distance profile for the code due to a
sub-optimal execution order of the operations (e.g., untiled code
version of a tileable algorithm) or is it more fundamental property of
the computation that remains relatively unchangeable through any
transformations that change the execution order of the operations?  This
is the question our work seeks to assist in answering. By analyzing
the execution trace of a given code, forming a dynamic data dependence
graph, and reordering the operations by forming convex partitions, the
potential for improving the reuse distance profile is
evaluated. The change to the reuse distance profile after the dynamic
analysis and reordering, rather than the shape of the initial reuse
distance profile of a code, provides guidance on the potential for
further improvement.

Fig.~\ref{fig:seidel1RDA}(c) presents the information in
Fig.~\ref{fig:seidel1RDA}(b) in terms of memory bandwidth
required per operation. 
It translates
the cache miss count into the bandwidth demand on the memory
system in bytes/second per floating-point operation.
For this code, we have one floating point operation per two memory
references. With a cache miss rate $m$, assuming double-precision (8 bytes per word),
the demand on the main-memory bandwidth would be $16*m$ bytes per Flop.
If this ratio is
significantly higher than the ratio of a system's main memory
bandwidth (in Gbytes/sec) to its peak performance (in GFlops), the
locality analysis indicates that achieving high performance will be
critically dependent on effective data locality optimization. For
example, on most current systems, the performance of this computation
will be severely constrained by main memory bandwidth for problem
sizes that are too large to fit in cache.

A point of note is that while the estimated cache hit rates/counts by
using RDA can deviate quite significantly from actually measured cache
hit rates/counts on real systems (due to a number of aspects of real
caches, such as non-unit line size, finite associativity,
pre-fetching, etc.), the bytes/flop metric serves as a good indicator
of the bandwidth requirement for real caches. This is because
pre-fetching and non-unit line sizes only affect the latency and
number of main memory accesses and not the minimum volume of data that
must be moved from memory to cache.  Finite cache associativity could
cause an increase in the number of misses compared to a fully
associative cache, but not a decrease.  All the experimental results
presented later in the article depict estimates on the bytes/flop
bandwidth demanded by a code.  Thus, despite the fact that RDA
essentially models an idealized fully associative cache, the data
represents useful estimates on the bandwidth demand for any real
cache.

\begin{figure}[h!tb]
\centering\begin{minipage}[b]{0.99\textwidth}
\begin{minipage}[b]{0.45\textwidth}
\centering\includegraphics[width=6cm]{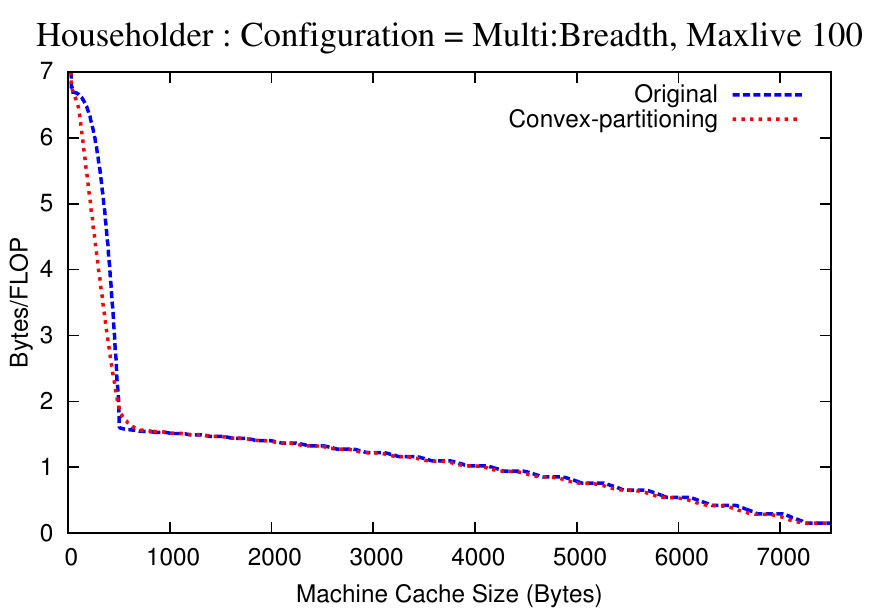}
\end{minipage}
~~~~~~
\begin{minipage}[b]{0.45\textwidth}
\centering\includegraphics[width=6cm]{./newfigs/outplace-fw_all_3n}
\end{minipage}
\end{minipage}
\vspace{-.2cm}
\caption{\label{fig:rdaIsCrap} Reuse distance analysis for Householder and Floyd-Warshall }
\end{figure}

\paragraph{Benefits of the proposed dynamic analysis}
Using results from two case studies presented later in the article, we
illustrate the benefits of the approach we develop.
Fig.~\ref{fig:rdaIsCrap} shows the original
reuse distance profiles
as well as the profiles after dynamic analysis and
convex partitioning,
for two benchmarks: Householder transformation
on the left, and Floyd-Warshall all-pairs shortest path on the right.
As seen in the left plot in Fig.~\ref{fig:rdaIsCrap}, with the
Householder code, no appreciable change to the reuse distance profile
results from the attempted reordering after dynamic analysis.  In
contrast, the right plot in Fig.~\ref{fig:rdaIsCrap} shows a
significantly improved reuse distance profile for the Floyd-Warshall
code, after dynamic analysis and reordering of operations. This
suggests potential for enhanced data locality via suitable code
transformations. As explained later in the experimental results
section, manual examination of the convex partitions provided insights
into how the code could be transformed into an equivalent form that in
turn could be tiled by an optimizing compiler. The reuse distance
profile of that tiled version is shown as a third curve in the right
plot in Fig.~\ref{fig:rdaIsCrap}, showing much better reuse than the
original code. The actual performance of the modified code was also
significantly higher than the original code. To the best of our
knowledge, this is the first 3D tiled implementation of the Floyd
Warshall algorithm (other blocked versions have been previously
developed \cite{sahni-FW,prasanna-FW}, but have required
domain-specific reasoning for semantic changes to form equivalent
algorithms that generate different intermediate values but the same
final results as the standard algorithm).

\vspace{-.4cm}
\vspace{-.7cm}

\subsection{Overview of Approach}
\label{sec:overview}

The new dynamic analysis approach proposed in this paper attempts to characterize
the inherent data locality properties of a given (sequential) computation, and
to assess the potential for enhancing data locality via change of
execution ordering. To achieve this goal, we proceed in two
stages. First, a new ordering of the program's operations is computed,
by using graph algorithms (that is, convex partitioning techniques)
operating on the expanded computation graph. Then, standard reuse
distance analysis is performed on the reordered set of operations. We
note that our analysis does not directly provide an optimized
program. Implementing the (possibly very complex) schedule found
through our graph analysis is impractical. Instead, our
analysis highlights gaps between the reuse distance profile
of a current implementation and
existing data locality potential: the task of devising a better
implementation is left to the user or compiler writer.
In Sec.~\ref{sec:experiments}, we show the benefits of the approach
on several benchmarks.

To implement our new dynamic analysis, we first analyze the data
accesses and dependences between the primitive operations in a
sequential execution trace of the program to extract a more abstract
model of the computation: a computational directed acyclic graph
(CDAG), where operations are represented as vertices and the flow of
values between operations as edges. This is defined as follows.
\begin{definition}[CDAG~\cite{bilardi2001characterization}]
A computation directed acyclic graph (CDAG) is a 4-tuple
$C = (I,V,E,O)$ of finite sets such that:
(1) $I \cap V = \emptyset$;
(2) $E \subseteq (I\cup V) \times V$ is the set of arcs;
(3) $G = (I\cup V, E)$ is a directed acyclic graph with no isolated vertices;
(4) $I$ is called the input set;
(5) $V$ is called the operation set and all its vertices have one
or two incoming arcs;
(6) $O \subseteq (I\cup V)$ is called the output set.
\end{definition}

\begin{figure}[h!tb]
\centering
\begin{minipage}{\textwidth}
\begin{minipage}{.45\textwidth}
\centering
\includegraphics[width=2.8cm]{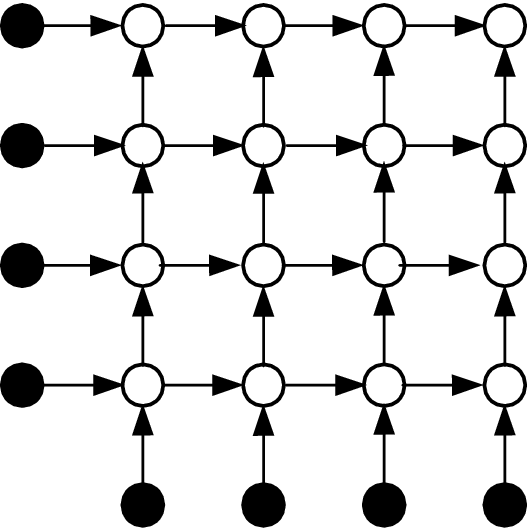}
\caption{\label{fig:Seidel-CDAG} CDAG for Gauss-Seidel code in
  Fig.~\ref{fig:seidel1code}. Input vertices are shown in black, 
  other vertices represent operations performed.}
\end{minipage}
\hfill
\begin{minipage}{.45\textwidth}
\centering
\includegraphics[width=3.3cm]{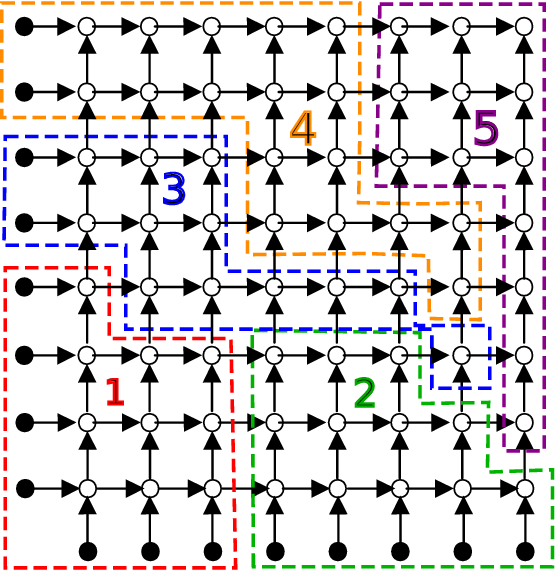}
\caption{\label{fig:heuristics-partition} Convex-partition of the
	CDAG for the code in Fig.~\ref{fig:seidel1code} for
	$N$ = 10.}
\end{minipage}
\end{minipage}
\end{figure}

Fig.~\ref{fig:Seidel-CDAG} shows the CDAG corresponding to the code in
Fig.~\ref{fig:seidel1code} for $N$=6 --- both versions have identical
CDAGs since they perform exactly the same set of floating-point
computations, with the same inter-instance data dependences, even
though the total order of execution of the statement instances is
different. The loop body performs only one addition and is executed a
total of 16 times, so the CDAG has 16 computation nodes (white
circles).

Although a CDAG is derived from analysis of dependences between
instances of statements executed by a sequential program, it abstracts
away that sequential schedule of operations and only imposes an
essential partial order captured by the data dependences between the
operation instances. Control dependences in the computation need not
be represented since the goal is to capture the inherent data locality
characteristics based on the set of operations that were actually
execution in the program.

They key idea behind the work presented in this article is to perform
analysis on the CDAG of a computation, attempting to find a different
order of execution of the operations that can improve the
reuse-distance profile compared to that of the given program's
sequential execution trace. If this analysis reveals a significantly
improved reuse distance profile, it suggests that suitable source code
transformations have the potential to enhance data locality. On the
other hand, if the analysis is unable to improve the reuse-distance
profile of the code, it is likely that it is already as well optimized
for data locality as possible.
The dynamic analysis involves the following steps:
\vspace{-.2cm}
\begin{enumerate}
\item Generate a sequential execution trace of a program.
\item Run a reuse-distance analysis of the original trace.
\item Form a CDAG from the execution trace.
\item Perform a multi-level convex partitioning of the CDAG, which
is then used to change the schedule of operations of the CDAG from
the original order in the given input code. A convex partitioning
of a CDAG is analogous to tiling the iteration space of a regular nested loop computation.
Multi-level convex partitioning
is analogous to multi-level cache-oblivious blocking.
\item Perform standard reuse distance analysis of the reordered trace
after multi-level convex partitioning.
\end{enumerate}

Finally, Fig.~\ref{fig:heuristics-partition} shows the convex
partitioning of the CDAG corresponding to the code in
Fig.~\ref{fig:seidel1code}.


After such a partitioning, the execution
order of the vertices is reordered so that the convex partitions are
executed in some valid order (corresponding to a topological sort of a
coarse-grained inter-partition dependence graph), with the vertices
within a partition being executed in the same relative order as the
original sequential execution. Details are presented in the next
section.



\section{Convex Partitioning of CDAG}
\label{sec:cp}

In this section, we provide details on our algorithm for convex
partitioning of CDAGs, which is at the heart of our proposed dynamic
analysis.  In the case of loops, numerous efforts have attempted to
optimize data locality by applying loop transformations, in particular
involving loop tiling and loop fusion
\cite{irigoin88popl,wolf.91.pldi,kennedy.93.lcpc,uday08pldi}.  Tiling
for locality attempts to group points in an iteration space of a loop
into smaller blocks (tiles) allowing reuse (thereby reducing reuse
distance) in multiple directions when the block fits in a faster
memory (registers, L1, or L2 cache).  Forming a valid tiling for a
loop requires that each tile can be executed atomically, i.e., each
tile can start after performing required synchronizations for the data
it needs, then execute all the iterations in the tile without
requiring intervening synchronization. This means that there are no
cyclic data dependencies between any two tiles.  \emph{Our goal in
  this work is to extend this notion of ``tiling'' to arbitrary CDAGs
  that represent a computation: we form valid partitioning of CDAGs
  into components such that the components can be scheduled and
  executed, with all vertices in a component being executed
  ``atomically,'' i.e., without being interleaved with vertices in any
  other components.}  For this, we rely on the notion of \emph{convex
  partitioning} of CDAGs, which is the generalization of loop
tiling to graphs.

\subsection{Definitions}

We first define what is a convex component, that is a tile in a graph.

\begin{definition}[Convex component]
\label{def:cp}
Given a CDAG $G$, a convex component $V_i$ in $G$ is defined as a
subset of the vertices of $G$ such that, for any pair of vertices
$u$ and $v$ in $V_i$, if there are paths between $u$ and $v$ in $G$,
then every vertex on every path between $u$ and $v$ also belongs to
$V_i$.
\end{definition}
\vspace{-.2cm}
A convex partition of a graph $G$ is obtained by assigning each vertex
of $G$ to a single convex component.  Since there are no cycles among
convex components, the graph in which nodes are convex components and
edges define dependences among them, is acyclic. Therefore, we can
execute the convex components using any topologically sorted order.
Executing all the convex components results in executing the full
computation.

A \emph{convex partition} of a graph $G=(V,E)$ is a collection of convex
components $\{V_1,\ldots,V_k\}$ of $G$ such that $\bigcup V_{i=1}^k =
V$ and for any $i, j$ s.t. $1 \leq i,j \leq k$ and $i \neq j$, $V_{i} \cap V_{j} = \emptyset$. 
We remark that tiling of iterations spaces of loops results in convex
partitions.

The \emph{component graph} $\mathcal{C} = (\mathcal{V_C},\mathcal{E_C})$
is defined as a graph whose vertices
$\mathcal{V_C}$ represent the convex components, i.e.,
$\mathcal{V_C} = \{V_1, \cdots, V_k\}$. Given two distinct components $V_i$
and $V_j$, there is an edge in $\mathcal{C}$ from $V_i$ to $V_j$ if
and only if there is an edge $(a,b)$ in the CDAG, where
$a \in V_i$ and $b \in V_j$.


For a given schedule of execution of the vertices of convex component $V_i$,
we define \emph{maxlive} to be the
maximum number of simultaneously live nodes for this schedule.
A node can be in one of the following states throughout its life:
\textit{(initial state)} at the beginning no node is live;
\textit{(birth)} any node is considered live right after it is fired (executed);
\textit{(resurrection)} if not part of the convex component, it is also considered
as live when used by another node of the component (predecessor of a fired node belonging to the component);
\textit{(live)} a born or resurrected node stays alive until it dies, which happens
if all its successor nodes have executed (are part of the partition);
\textit{(death)} a node dies right after its last successor fires.

Our goal is to form convex components along with a scheduling such that
the \emph{maxlive} of each component does not exceed the local memory capacity.
We consider the nodes we add to the component (just fired and alive),
and their predecessors (resurrected) in computing the \emph{maxlive}.



\subsection{Forming Convex Partitions}

We show in Algorithm~\ref{alg:core-alg} our technique to build convex
partitions from an arbitrary CDAG. It implements a convex-component
growing heuristic that successively adds ready vertices into the
component until a capacity constraint is exceeded. The key requirement
in adding a new vertex to a convex component is that if any path to
that vertex exists from a vertex in the component, then all vertices
in that path must also be included. We avoid an expensive search for
such paths by constraining the added vertices to be those that already
have all their predecessors in the current (or previously formed)
convex component.

\vspace{-.4cm}
\begin{algorithm}[h]
\DontPrintSemicolon
\SetInd{0.5em}{0.5em}
\SetKwInOut{Input}{Input}
\SetKwInOut{InOut}{InOut}
\SetKwFunction{ready}{getTheInitialReadyNodes}
\SetKwFunction{getnode}{selectReadyNode}
\SetKwFunction{bestnode}{selectBestNode}
\SetKwFunction{updateR}{UpdateListOfReadyNodes}
\SetKwFunction{readyS}{findNeighborsAndSuccessors}
\SetKwFunction{maxlive}{Liveset}
\SetKwFunction{CF}{CF}
\SetKwFunction{outputset}{Outputset}
\SetKwFunction{getScn}{getSpecifiedNoOfNodesFromList}
\SetKwFunction{getCcn}{getSpecifiedNoOfNodesFromList}
\SetKwFunction{pushvec}{pushVector}
\SetKwFunction{updatelive}{updateLiveSet}
\SetKwFunction{newCC}{newConvexComponent}
\Input{
  $G$ : CDAG;  $C$ : Cache Size; \\
  \hspace{3pt} \textit{CF}: Cost function to decide next best node
}
\InOut{  $P$ : Partition containing convex components}

\Begin
  {
    $P \longleftarrow \emptyset$\;
    $R	\longleftarrow $  \ready{$G$} \;
    
    \While{ $R \neq \emptyset $} {

      $n \longleftarrow $	\getnode{$R$} \;	
      $cc \longleftarrow \emptyset$\;
      \While{$R\neq \emptyset \wedge \updatelive{\textit{cc, n, C}}$}{
        $cc \longleftarrow cc \cup \{n\}$ \;
        $R \longleftarrow R - \{n\}$\;
        \updateR{$R$, $n$}\;
	$priority \longleftarrow \CF{}$ \;
        $n \longleftarrow$ \bestnode{\textit{R, cc, priority, n}}\;
      }
    $P \longleftarrow P \cup \{cc\}$ 
    }
  }
  \caption{GenerateConvexComponents(\textit{G, C, CF}) \label{alg:core-alg}}
\end{algorithm}


The partitioning heuristic generates a valid schedule as it
proceeds. At the beginning, all input vertices to the CDAG are placed
in a ready list $R$.  A vertex is said to be \emph{ready} if all its
predecessors (if any) have already executed, i.e., have been assigned
to some convex component.  A new convex component $cc$ is started by
adding a ready vertex to it (the function
\texttt{selectReadyNode}($R$) simply picks up one element of $R$) and
it grows by successively adding more ready nodes to it
(\texttt{selectBestNode}(\textit{R, cc, priority, n}) selects one of the ready
nodes, as shown in Algorithm~\ref{alg:cost1} -- the criterion is
described later).  Suppose a vertex $n$ is just added to a component
$cc$. As a result, zero or more of the successors of $n$ in $G$ may
become ready: a successor $s$ of $n$ becomes ready if the last
predecessor needed to execute $s$ is $n$.
The addition of newly readied vertices to the ready list is done by
the function \texttt{updateListOfReadyNodes}(\textit{R, n}), as shown
in Algorithm~\ref{alg:updateready}.  In this function, the test that
checks if $s$ has unprocessed predecessors is implemented using a
counter that is updated whenever a node is processed.
\vspace{-.4cm}
\begin{algorithm}[h!tb]
\DontPrintSemicolon
\SetInd{0.5em}{0.5em}
\SetKwInOut{Input}{Input}
\SetKwInOut{InOut}{InOut}
\SetKwFunction{child}{successors}
\Input{$n$: Latest processed node}
\InOut{$R$: List of ready nodes}
\Begin
  {
    \For{$s \in $ \child{$n$}}
        {
          \If{$s$ has no more unprocessed predecessors}
             {
               $R \longleftarrow R \cup \{s\}$ \;
             }
        }
  }
  \caption{\label{alg:updateready}UpdateListOfReadyNodes(\textit{R, n})}
\end{algorithm}

\vspace{-.45cm}

Before adding a node to $cc$, the set \textit{cc.liveset}, the liveout
set of $cc$, is updated through the call to
\texttt{updateLiveSet}\textit{(p, n, C)}, as shown in
Algorithm~\ref{alg:updatelive}. \texttt{updateLiveSet} exactly
implements our definition of liveness previously described:
\textit{(birth)} if $n$ has some successors it is added to the liveset
of $cc$; \textit{(resurrect)} its predecessor nodes that still have
unprocessed successors are added to the liveset (if not already in
it); \textit{(die)} predecessor nodes for which $n$ is the last
unprocessed successor are removed from the liveset.
\vspace{-.4cm}
\begin{algorithm}[h!tb]
\DontPrintSemicolon
\SetInd{0.5em}{0.5em}
\SetKwInOut{Input}{Input}
\SetKwInOut{Output}{Output}
\SetKwInOut{InOut}{InOut}
\SetKwFunction{getpred}{predecessors}
\SetKwFunction{size}{size}
\SetKwFunction{getbase}{FirstLevelBaseNodes}
\SetKwFunction{remove}{removeNodeFromSet}

\Input{
  $n$ : New node added in the partition $p$ \\
  ~~~$C$ : Cache size
}
\InOut{
  $p.\textit{liveset}$ : Live set of $p$
}
\Output{
  true if $|p.liveset| \leq C$, false otherwise
}
\Begin{
      $lset \longleftarrow p.liveset$\;
      \If{$n$ has unprocessed successors}{
        $p.\textit{liveset} \longleftarrow p.\textit{liveset} \cup \{n\}$\;
      }
      \For{$n' \in \getpred{\textit{n}} $}{
        \If{$n'$ has unprocessed successors}{
          $p.\textit{liveset} \longleftarrow p.\textit{liveset} \cup \{n'\}$\;
        }
        \ElseIf{$n' \in p.\textit{liveset}$}{
          $p.\textit{liveset} \longleftarrow p.\textit{liveset}-\{n'\}$}
      }
      \If{$|p.liveset| > C$}{
         $p.liveset \longleftarrow lset$\;
         \Return false\;
      }
      \Return true\;
  }

\caption{\label{alg:updatelive}updateLiveSet(\textit{p, n, C})}
\end{algorithm}

\vspace{-.7cm}

\subsection{CDAG Traversal: Breadth-first Versus Depth-first}

The heuristic \texttt{selectBestNode} to select the next processed node within the ready list
uses two affinity notions: a node is a ready-successor of $cc$ (thus
element of the \textit{cc.readySuccessors} list) if it is a ready
successor of some nodes of $cc$; a node is a ready-neighbor of $cc$
(thus element of \textit{cc.readyNeighbors} list) if it has a
successor node that is also a successor of some node in $cc$. We note
that those two lists can overlap. The growing strategy picks up ready
nodes, using a first-in first-out policy, from one of those two
lists. In practice, we observe that favoring nodes of the
ready-successor list would favor growing depth-first in the CDAG,
while favoring nodes that belongs to the ready-neighbor list would
favor a breadth-first traversal.  

The heuristic uses a combination of growing alternately in these two
different directions till the $Maxlive$ of the component exceeds the
capacity of the cache. The priority is controlled by a number that
represents the ratio of selected ready-neighbors over selected
ready-successors. If the ratio is larger than $1$, we refer to it as
\emph{breadth-priority}; if lower than $1$, we refer to it as
\emph{depth-priority}, otherwise it is referred to as
\emph{equal-priority}. Function $neighbor(n)$ will return the list of
all nodes (excluding $n$) that have a common successor with $n$.
\vspace{-.45cm}
\begin{algorithm}[h!tb]
\DontPrintSemicolon
\SetInd{0.5em}{0.5em}
\SetKwInOut{Input}{Input}
\SetKwInOut{Output}{Output}
\SetKwInOut{InOut}{InOut}
\SetKwFunction{first}{dequeue}
\SetKwFunction{push}{enqueue}
\SetKwFunction{getN}{findAllReadyNeighbors}
\SetKwFunction{getS}{findAllReadySuccessors}
\Input{
  $R$: List of ready nodes\\ 
  \hspace{5pt} $cc$: current convex component\\
  \hspace{5pt} $priority$: Give priority to neighbor or successor\\
  \hspace{5pt} $n$: Latest node added in the partition \\
}
\InOut{
  \textit{cc.readyNeighbors} : Ready neighbors of current growing partition nodes \\
  \hspace{8pt}\textit{cc.readySuccessors} : Ready successors of current growing partition nodes
}
\Output{
  $next$ : Next node to add in the partition 
}
\Begin
  {
    \For{$a \in \texttt{neighbors}(n)\cap R - \textit{cc.readyNeighbors}$}{
      \textit{cc.readyNeighbors}.\push{$a$}\;
    }
    \For{$a \in \texttt{successors}(n)\cap R - \textit{cc.readySuccessors}$}{
      \textit{cc.readySuccessors}.\push{$a$}\;
    }
    $\textit{cc.readyNeighbors} \longleftarrow \textit{cc.readyNeighbors} - n$ \;
    $\textit{cc.readySuccessors} \longleftarrow \textit{cc.readySuccessors} - n$ \;
    
    \If{$\textit{cc.takenNeighbors} < \textit{cc.takenSuccessors}\times \textit{priority}$\\ \hspace{1cm}$\wedge \ \textit{cc.readyNeighbors}\neq\emptyset$}
    {
      $\textit{next} \longleftarrow \first{\textit{cc.readyNeighbors}}$ \;
      $\textit{cc.takenNeighbors} \longleftarrow \textit{cc.takenNeighbors}+1$\;
    }
    \ElseIf{$\textit{cc.readySuccessors}\neq\emptyset$}
    {
      $\textit{next} \longleftarrow \first{\textit{cc.readySuccessors}}$ \;
      $\textit{cc.takenSuccessors} \longleftarrow \textit{cc.takenSuccessors}+1$\;
    }
    \Else{
      $\textit{next} \longleftarrow \texttt{selectReadyNode}(R)$\;
    }

    \Return \textit{next} \;
  }
\caption{selectBestNode(\textit{R, cc, priority, n})}
\label{alg:cost1}
\end{algorithm}

\vspace{-.5cm}


\subsection{Multi-level Cache-oblivious Partitioning}

Here we address the problem of finding a schedule that is cache-size
oblivious, and in particular suitable for multi-level memory
hierarchy. In order to address this problem, we construct a
hierarchical partitioning using the multi-level component growing
heuristic shown in Algorithm~\ref{alg:multilevel}. This algorithm
combines individual components formed for a cache of size $C$ using
the heuristic in Algorithm~\ref{alg:core-alg} to form components for a
cache of size $\textrm{factor * }C$. In this approach, each component
of the partition built at level $l$ of the heuristic is seen as a
single node at level $l+1$. We call these nodes ``macro-nodes'' as
they typically represent sets of nodes in the original CDAG.

This approach could be compared with
multi-level tiling for multi-level cache hierarchy, a classical scheme
to optimize data locality for multi-level caches.
For the first level of this heuristic, a macro-node corresponds to a
node in the original CDAG.  The heuristic then proceeds with the next
level, seeing each component of the partition at the previous level as
a macro-node at the current level. The heuristic stops when only one
component is generated at the current level, that is, all macro-nodes
were successfully added to a single convex component without exceeding
the input/output set size constraints.
%
The number of levels in the multi-level partitioning varies with each
CDAG, and is not controlled by the user.  When a component $cc_0$
formed at a lower level is added to the current component $cc_1$ being
formed at a higher level, the \textit{liveset} of $cc_1$ has to be
updated, just as if all the nodes composing $cc_0$ have been added to
$cc_1$. This leads to the modified version of
\texttt{updateLiveSet}(\textit{p,n,C}) reported in
Algorithm~\ref{alg:updatelive:multi}, where the function
\texttt{FirstLevelBaseNodes}(\textit{np}) returns the actual CDAG
nodes (that we call first level base nodes) the macro-node
\textit{np} is  built upon. At the first level
\texttt{FirstLevelBaseNodes}(\textit{np}) may be understood as
returning just \textit{np}.
 \vspace{-.5cm}

\begin{algorithm}[h]
\DontPrintSemicolon
\SetInd{0.5em}{0.5em}
\SetKwInOut{InOut}{InOut}
\SetKwInOut{Input}{Input}
\SetKwInOut{Output}{Output}
\SetKwFunction{size}{size}
\SetKwFunction{formmacro}{formMacroNodeWithEachPartition}
\SetKwFunction{partition}{GenerateConvexComponents}
\Input{
  $G$ : CDAG\\
  $C$ : initial cache Size\\
  \textit{Priority} : priority to Neighbor or Successor \\
  \textit{factor} : multiplication factor of Cache size for each level
}
\InOut{
  $G.M$ : memory footprint of the CDAG
}
\Output{  $P$ : Final single partition}

\Begin{

 $P \longleftarrow$ \partition{\textit{G, C, Priority}}\;
    
     \While{$C < G.M$}
    {  
       $G' \longleftarrow$ \formmacro{$P$, $G$}\;
        $C \longleftarrow factor*C$\;
        $P' \longleftarrow$ \partition{$G'$, $C$, $Priority$}\;
        $P \longleftarrow P'$   \;
    }
    \Return $P$ \;

}
\caption{MultiLevelPartitioning(\textit{G, C, Priority, factor}) \label{alg:multilevel}}
\end{algorithm}

 \vspace{-1cm}
\begin{algorithm}[h]
\DontPrintSemicolon
\SetInd{0.5em}{0.5em}
\SetKwInOut{Input}{Input}
\SetKwInOut{Output}{Output}
\SetKwInOut{InOut}{InOut}
\SetKwFunction{getpred}{predecessors}
\SetKwFunction{size}{size}
\SetKwFunction{getbase}{FirstLevelBaseNodes}
\SetKwFunction{remove}{removeNodeFromSet}

\Input{
  $n$ : New macro-node added in the partition $p$
}
\InOut{
  $p.\textit{liveset}$ : Live set of $p$
}
\Output{
  true if $|p.liveset| \leq C$, false otherwise
}
\Begin{
    $b \longleftarrow$ true\;
    $plset \longleftarrow p.liveset$\;
    \For{$nb \in \getbase{n}$}
    {
      $b \longleftarrow b~\land$ \texttt{updateLiveSet}(\textit{p, nb, C})
    }
    \If {$b = $~false}{
       $p.liveset \longleftarrow plset$\;
       \Return false\;
    }
    \Return true\;
  }

\caption{\label{alg:updatelive:multi}updateLiveSet(\textit{p, n, C})}
\end{algorithm}

 \vspace{-.4cm}

\paragraph{Complexity Analysis}
The overall complexity of the single-level version is thus linear with
the size of the trace.  The multi-level version is run
$\log_{factor}\left(\frac{|M|}{|C|}\right)$ times
\texttt{GenerateConvexComponents}, thus leading to an overall time
complexity of $O(|T|\log(|M|))$. A step-by-step analysis of the complexity of our analysis can be found in \cite{taco14TR}.

\section{Experimental Results}
\label{sec:experiments}

The experimental results are organized in two major parts. In
Sec.~\ref{subsec:heuristcomp} we evaluate the impact of the various
parameters of the convex
partitioning heuristics, using two well understood benchmarks: matrix
multiplication and a 2D Jacobi stencil computation. With these
benchmarks, it is well known how the data locality characteristics
of the computations can be improved via loop tiling. The goal therefore
is to assess how the reuse distance profiles after dynamic analysis
and operation reordering compares with the unoptimized untiled original code
as well as the optimized tiled version of code.

In Sec.~\ref{subsec:casestudies} we detail several case studies where
we use dynamic analysis to characterize the locality potential of
several benchmarks for which the state-of-the-art optimizing compilers
are unable to automatically optimize data locality.  We demonstrate
the benefits of dynamic analysis in providing insights into the
inherent data locality properties of these computations and the
potential for data locality enhancement. Finally we discuss in
Sec.~\ref{subsec:datasets} the sensitivity of our techniques to
varying datasets.

\subsection{Experimental Setup}

The dynamic analysis we have implemented involves three steps. For the
CDAG Generation, we use automated LLVM-based instrumentation to
generate the sequential execution trace of a program, which is then
processed to generate the CDAG.  The trace generator was previously
developed for performing dynamic analysis to assess vectorization
potential in applications \cite{holewinski-pldi2012}. For the convex
partitioning of the CDAG, we have implemented the algorithms explained
in detail in the previous section. Finally for the reuse distance
analysis of the reordered address trace after convex partitioning, it
is done using a parallel reuse distance analyzer PARDA \cite{parda}
that was previously developed.

The total time taken to perform the dynamic analysis is dependent on
the input program trace size. In our experiments, computing the trace
and performing the full dynamic analysis can range between seconds for
benchmarks like Givens, Householder, odd-even sort or Floyd-Warshall
to about one hour for SPEC benchmarks such as 420.LBM. For instance,
for Givens rotation (QR decomposition) the trace is built in 4
seconds, the analysis takes another 14 seconds, and computing the
reuse distance analysis on the partitioned graph takes well below a
second. We note that while CDAGs are often program-dependent, the
shape of the CDAG and the associated dynamic analysis reasoning can
usually be performed on a smaller problem size: the conclusion about
the data locality potential is likely to hold for the same program
running on larger datasets. A study of the impact of the sensitivity
of our analysis to different datasets is provided in later
Sec.~\ref{subsec:datasets}. All performance experiments were done on
an Intel Core i7 2700K, using a single core.



\subsection{Impact of Heuristic Parameters}
\label{subsec:heuristcomp}

The convex partitioning heuristic takes two parameters. First the {\em
  Search Strategy}: this includes (a) prioritization in selecting a
new vertex to include in a convex partition: depth-priority,
breadth-priority, or alternation between depth and breadth priority
(equal priority); and (b) single level partitioning versus multi-level
partitioning. The second is {\em Maxlive}: the parameter that sets a
limit on the maximum number of live vertices allowed while forming a
partition.

\subsubsection{Jacobi 2D}

\begin{figure}[h!tb]
  \centering
  \mbox{
    \subfloat{
      \begin{minipage}[b]{0.45\textwidth}
      \includegraphics[width=6.7cm]{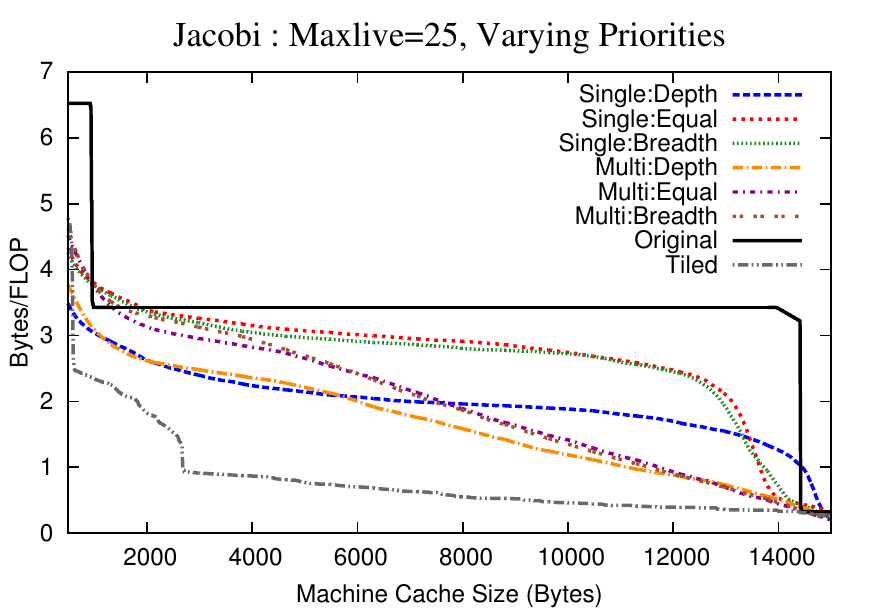}
      \label{fig:jacobi_priorities}
      \end{minipage}
    }
    \hspace{1cm}
   \subfloat{
      \begin{minipage}[b]{0.45\textwidth}
      \includegraphics[width=6.7cm]{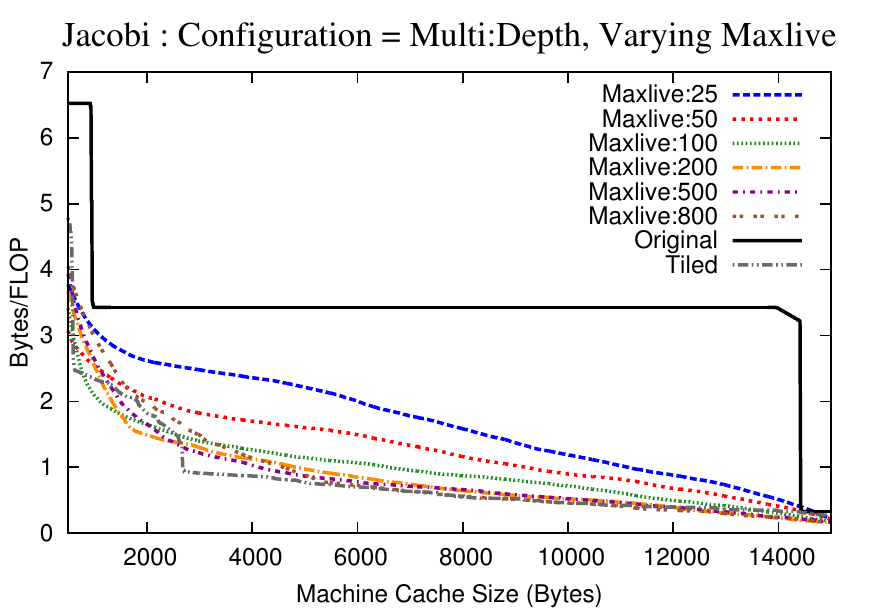}
      \label{fig:jacobi_cache_variation}
      \end{minipage}
    }
 }
\vspace{-.7cm}
\caption{Results with different heuristics for Jacobi-2D}
\vspace{-.3cm}
\label{fig:jacobi}
\end{figure}

Fig.~\ref{fig:jacobi} presents the results of applying the dynamic
analysis on a Jacobi stencil on a regular 2-dimensional grid of size
32, and 30 time iterations.

\vspace{-.45cm}

Fig.~\ref{fig:jacobi}-left shows reuse
distance profiles for a fixed value of $Maxlive$ and different
configurations for single versus multi-level partitioning, and
different priorities for next node selection.  With single-level
partitioning, depth-priority is seen to provide the best
results. Using multi-level partitioning further improves the reuse
distance profile. In order to judge the effectiveness of the convex
partitioning in improving the reuse distance profile, we show both the
reuse distance profiles for the original code and an optimally tiled
version of the Jacobi code. It can be seen that there is significant
improvement over the original code, but still quite some distance from
the profile for the tiled code.  Fig.~\ref{fig:jacobi}-right shows the
effect of varying $Maxlive$ from 25 to 800, with multi-level
depth-priority partitioning.

Here it can be seen that at large values
of $Maxlive$, the profile is very close to that of the optimized tiled
code. Thus, with a large value of $Maxlive$ and use of multi-level
depth-priority partitioning, the convex partitioning heuristic is very
effective for the Jacobi-2D benchmark.

\subsubsection{Matrix Multiplication}

\begin{figure}[h!tb]
  \centering
  \mbox{
    \subfloat{
      \begin{minipage}[b]{0.45\textwidth}
      \includegraphics[width=6.7cm]{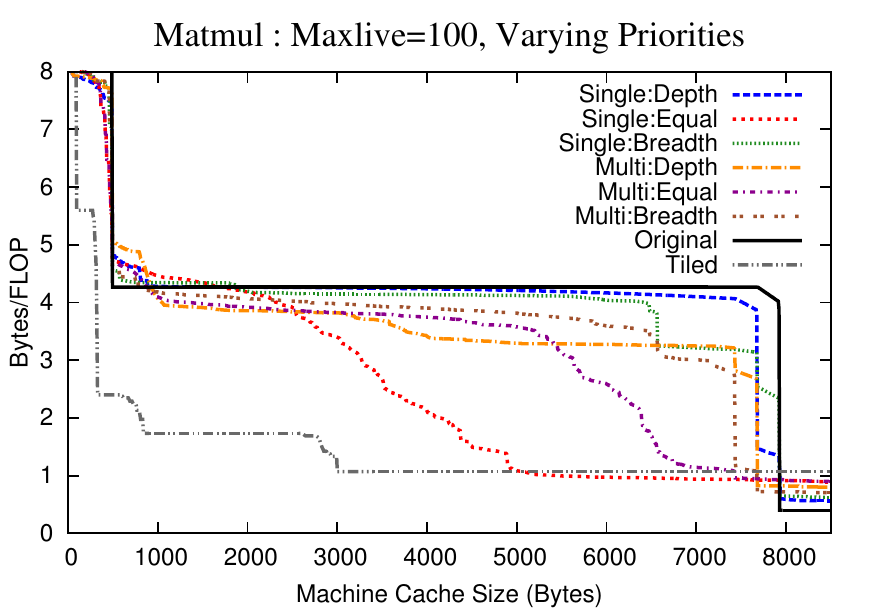}
      \label{fig:matmul_priorities}
      \end{minipage}
    }
    \hspace{0.7cm}
   \subfloat{
      \begin{minipage}[b]{0.45\textwidth}
      \includegraphics[width=6.7cm]{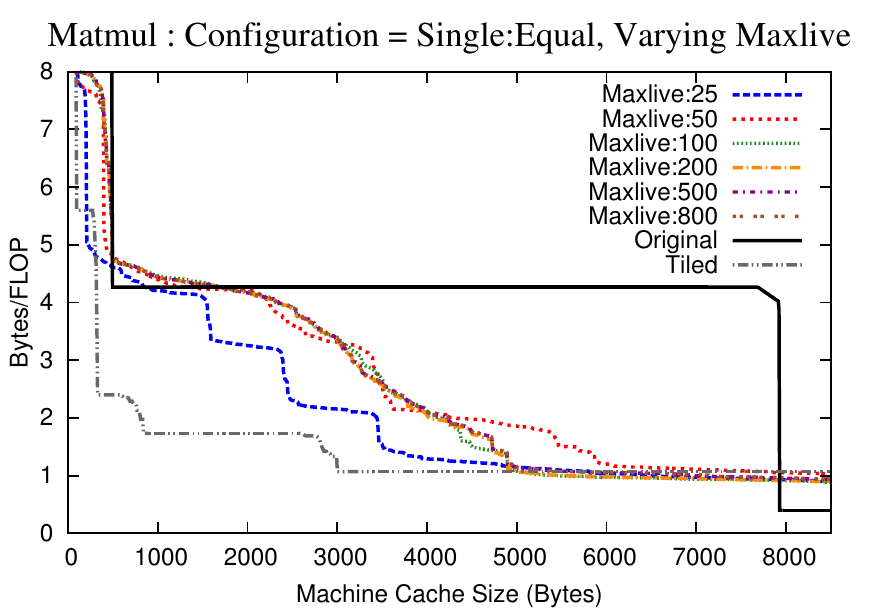}
      \label{fig:matmul_cache_variation}
      \end{minipage}
    }
 }
\vspace{-.45cm}
\caption{Results with different heuristics for matrix multiplication}
\vspace{-.4cm}
\label{fig:matmul}
\end{figure}

Fig.~\ref{fig:matmul} shows experimental results for matrix
multiplication, for matrices of size 30 by 30.  In
Fig.~\ref{fig:matmul}-left, the selection priority is varied, for single
and multi-level partitioning. In contrast to the Jacobi benchmark, for
Matmult, equal priority works better than breadth or depth priority.
Further, single level partitioning provides better results than
multi-level partitioning. In Fig.~\ref{fig:matmul}-right, we see
performance variation as a function of $Maxlive$ for single-level
equal-priority partitioning. Again the trends are quite different from
those of Jacobi-2D: the best results are obtained with the lowest
value of 25 for $Maxlive$.
\vspace{-.35cm}

These results suggest that no single setting of parameters
for the convex partitioning heuristic is likely to be consistently
effective across benchmarks.
We conjecture that there may be a relationship between
graph properties of the CDAGs (e.g., low fan-out vs. high
fan-out) and the best parameters for the partitioning heuristic.

profile than the best reported heuristic in
Fig.~\ref{fig:matmul}. This may happen when the tiled implementation
achieves the provably optimal I/O lower bound, which is the case for
Matmult here. Our heuristics for building convex partitions use
several simplifications to improve scalability, in particular in the
choice of candidates to be inserted in a partition, and in scheduling
the obtained partitions. In addition, we do not explore the Cartesian
product of all possible parameters values (priority and maxlive
values) but instead limit to a reasonable subset for scalability
purposes. All these factors contribute to our analysis possibly
under-estimating the data locality potential of the application. We
discuss in Sec.~\ref{sec:discussion} the tightness of our analysis and
what complementary techniques could be used to assess the quality of
our heuristics.

\subsection{Case Studies}
\label{subsec:casestudies}

We next present experimental results from applying dynamic analysis to
several benchmarks: the Floyd-Warshall algorithm to find all-pairs
shortest path in a graph represented with an adjacency matrix, two QR
decomposition methods: the Givens rotation and the Householder
transformation, three SPEC benchmarks, a LU decomposition code from
the LAPACK package, and an implementation of odd-even sorting using
linked list.  None of these benchmarks could be fully tiled for
enhanced data locality by state-of-the art research compilers (e.g.,
Pluto~\cite{pluto}) or by production compilers (e.g., Gnu GCC, Intel
ICC).  For each benchmark, we study the reuse distance profile of the
original code and the code after convex partitioning. Where
significant potential for data locality enhancement was revealed by
the dynamic analysis, we analyzed the code in greater detail. For two
of the four benchmarks we manually optimized (namely, Floyd-Warshall
and Givens rotation) have static control-flow. Therefore, the
performance for these will be only a function of the dataset size and
not the content of the dataset. For all optimized benchmarks, we
report performance on various dataset sizes (and various datasets when
relevant).

\subsubsection{Floyd-Warshall}
\label{sec:cs-fw}

\paragraph{Original program} We show in Fig.~\ref{lst:fw-orig} the original
input code that we used to implement the Floyd-Warshall algorithm. We
refer to this code as ``out-of-place'' Floyd-Warshall because it uses
a temporary array to implement the all-pairs shortest path
computation.

\begin{figure}[h!tb]
\vspace{-.3cm}
\centering\begin{minipage}{.7\textwidth}
{\scriptsize
\begin{lstlisting}[basicstyle=\scriptsize,frame=single]
for (k = 0; k < N; k++) {
  for (i = 0; i < N; i++)
    for (j = 0; j < N; j++)
      temp[i][j] = MIN(A[i][j], (A[i][k] + A[k][j]));
  k++;
  for (i = 0; i < N; i++)
    for (j = 0; j < N; j++)
      A[i][j] = MIN(temp[i][j], (temp[i][k] + temp[k][j]));
}
\end{lstlisting}
}
\end{minipage}
\vspace{-.2cm}
\caption{\label{lst:fw-orig}Floyd-Warshall all-pairs shortest path}
\vspace{-.5cm}
\end{figure}

\paragraph{Analysis}

\begin{figure}[h!tb]
\vspace{-.3cm}
  \centering
  \mbox{
    \subfloat{
      \begin{minipage}[b]{0.45\textwidth}
      \includegraphics[width=6.7cm]{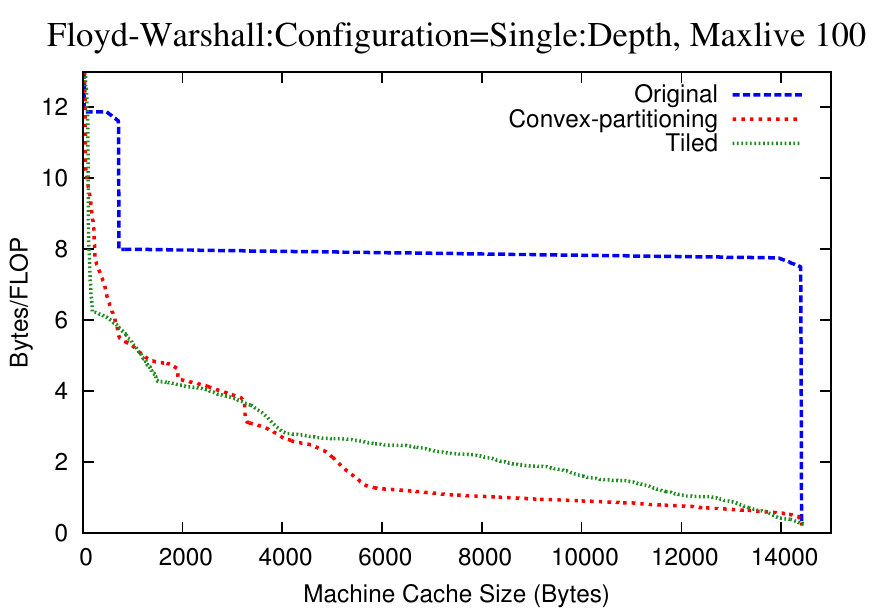}
      \label{fig:outplace-fw-all-3}
      \end{minipage}
    }
    \hspace{0.5cm}
   \subfloat{
      \begin{minipage}[b]{0.45\textwidth}
      \includegraphics[width=6.7cm]{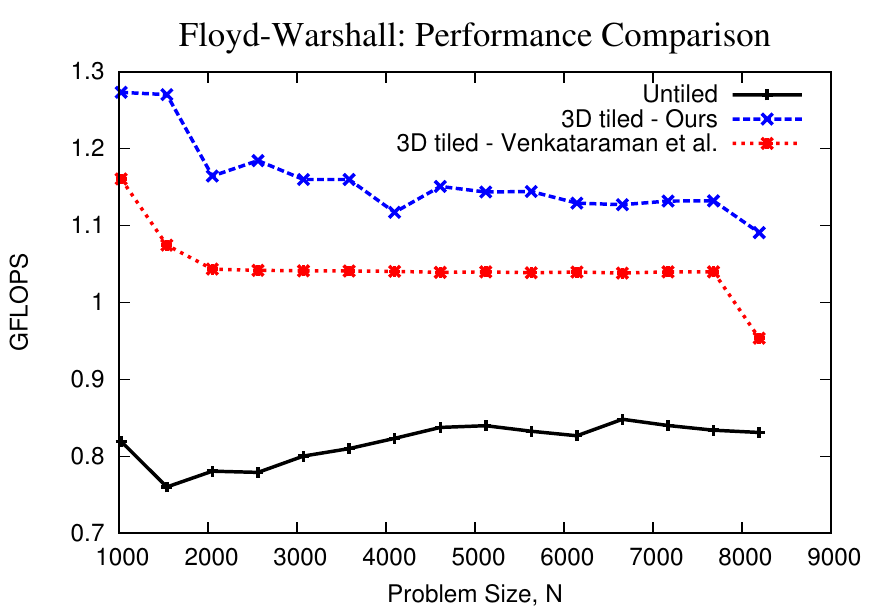}
      \label{fig:outplace-fw-perf}
      \end{minipage}
    }
 }
\vspace{-.45cm}
\caption{Floyd-Warshall: Analysis and performance improvements due to tiling
    \label{fig:fw-comparison}}
\vspace{-.4cm}
\end{figure}

Fig.~\ref{fig:fw-comparison}-left shows the reuse distance profile of
the original code, and the best convex-partitioning found by our
heuristics for the Floyd-Warshall algorithm, for a matrix of size 30
by 30. Since the convex-partitioning heuristic shows significantly
better reuse distance profile than the original code, there is
potential for improvement of data locality through transformations for
this program. Studies on the impact of the various heuristic
parameters on the quality of the convex partitioning obtained can be
found in \cite{taco14TR}.




Indeed, the Floyd-Warshall algorithm is immediately tilable, along the
two loops \texttt{i} and \texttt{j}. Such tiling can for instance be
achieved automatically with polyhedral model based compilers such as
Pluto \cite{uday08pldi}. Since the three loops are not fully permutable,
it has been believed that the Floyd-Warshall code cannot be 3D-tiled
without transformations using semantic properties of the algorithm
to create a modified algorithm (i.e., with a different CDAG) that provably
produces the same final result \cite{sahni-FW,prasanna-FW}.
However, a careful inspection of the convex
partitions revealed that \emph{valid 3D tiles can be formed among
the operations of the standard Floyd-Warshall algorithm}. This non-intuitive result comes from the
non-rectangular shape of the tiles needed, with varying tile size
along the \texttt{k} dimension as a function of the value of
\texttt{k}. This motivated us to look for possible transformations
that could enable 3D tiling of the code without any semantic transformations.

\paragraph{Modified implementation}
The modified implementation we designed can be found
in \cite{taco14TR}. It has an identical CDAG, i.e., it is semantically
equivalent, to the code in Listing~\ref{lst:fw-orig}. To create this
version, we fist split the original iteration space into four distinct
regions through manual index splitting, followed by tiling of each
loop nest.

\paragraph{Performance comparison}
Fig.~\ref{fig:fw-comparison} compares the performance of the tiled
version against the original code.
%
%
%
%
From the left plot in Fig.~\ref{fig:fw-comparison}, we can observe that the tiled
code is able to achieve better data locality, that is close to the
potential uncovered by the convex partitioning heuristics.
The right plot in Fig.~\ref{fig:fw-comparison} shows the improvement in actual
performance of our tiled code (\textsf{3D tiled - Ours}), due to
reduced cache misses.  A performance improvement of about 1.6$\times$
(sequential execution) is achieved, across a range of problem
sizes. Further, to the best of our knowledge, this is the first
development of a tiled version of the standard Floyd-Warshall code
that preserves the original code's CDAG.  We also compared with the performance
achieved by the semantically modified 3D-tiled implementation
from~\cite{sahni-FW} and found it to have slightly lower performance.

\subsubsection{Givens Rotation}
\label{sec:cs-givens}
\paragraph{Original program} Fig.~\ref{lst:givens-orig} shows the original
input code for the Givens rotation method used for QR decomposition.

\begin{figure}[h!tb]
\vspace{-.3cm}
\centering\begin{minipage}{0.8\textwidth}
{\scriptsize
\begin{lstlisting}[basicstyle=\scriptsize,frame=single]
for (j = 0; j < N; j++) {
  for (i = M-2; i >= j; i--) {
    double c =  A[i][j] / sqrt(A[i][j]*A[i][j] + A[i+1][j]*A[i+1][j]);
    double s = -A[i+1][j] / sqrt(A[i][j]*A[i][j] + A[i+1][j]*A[i+1][j]);
    for (k = j; k < N; k++) {
      double t1 = c * A[i][k] - s * A[i+1][k];
      double t2 = s * A[i][k] + c * A[i+1][k];
      A[i][k]   = t1;
      A[i+1][k] = t2;
    }
  }
}
\end{lstlisting}
}
\end{minipage}
\vspace{-.3cm}
\caption{\label{lst:givens-orig} Givens Rotation}
\vspace{-.6cm}
\end{figure}

\paragraph{Analysis}

\begin{figure}[h!tb]
\centering
\mbox{
	\subfloat{
		\begin{minipage}[b]{0.45\textwidth}
		\includegraphics[scale=0.8]{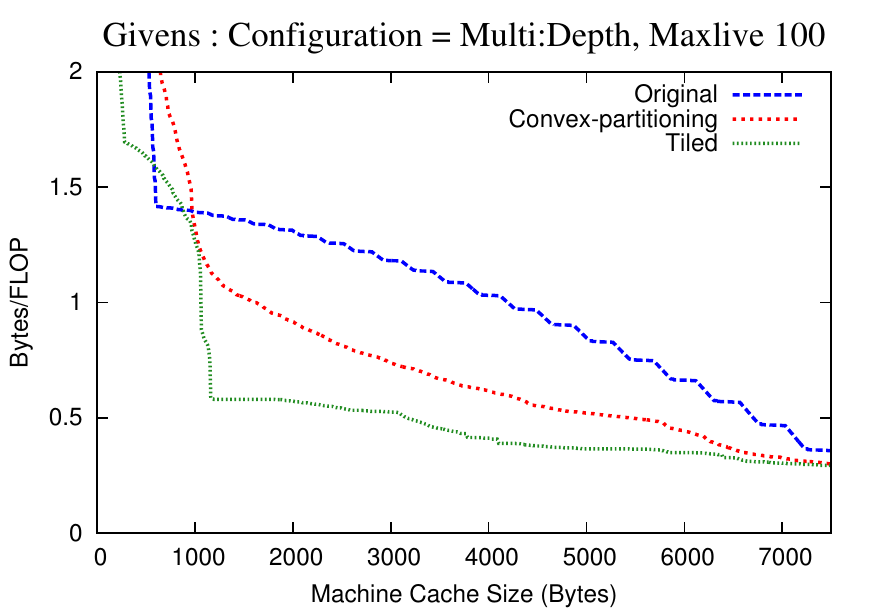}
		\label{fig:givens-reuse}
			\end{minipage}
	}
	\hspace{0.5cm}
	\subfloat{
		\begin{minipage}[b]{0.45\textwidth}
		\includegraphics[scale=0.8]{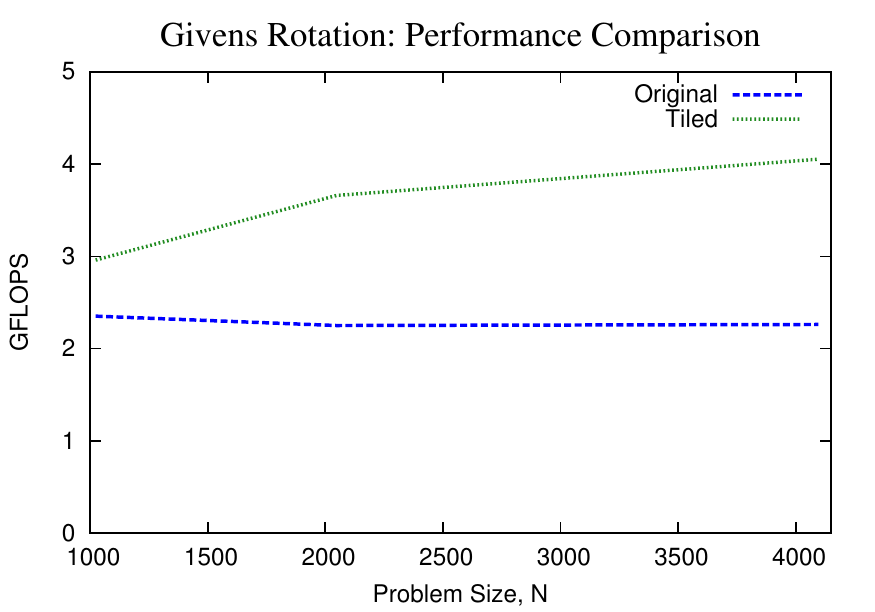}
		\label{fig:givens-perf}
			\end{minipage}
	}
}
\vspace{-.4cm}
\caption{Givens Rotation: performance improvements due to tiling
	\label{fig:givens-comparison}}
\vspace{-.7cm}
\end{figure}

Fig.~\ref{fig:givens-comparison} shows the reuse distance profile of
the original code and after convex-partitioning for the Givens
rotation algorithm, for an input matrix of size 30 by 30. Studies on the
impact of the various heuristic parameters on the quality of the
convex partitioning obtained can be found in \cite{taco14TR}.



The convex partitioning analysis shows good potential
for data locality improvement. Similarly to Floyd-Warshall, this code
can be automatically tiled by a polyhedral-based compiler \cite{uday08pldi},
after implementing simple loop normalization techniques. However, this
is not sufficient to tile all dimensions. Subsequent transformations
are needed, as shown below.

\paragraph{Modified implementation}

Based on the indicated potential for data locality enhancement, the
code in Listing~\ref{lst:givens-orig} was carefully analyzed and then
manually modified to enhance the applicability of automated tiling
techniques. Fig.~\ref{lst:givens-mod} shows this modified version. 

\begin{figure}[h!tb]
\vspace{-.3cm}
\centering\begin{minipage}{0.95\textwidth}
{\scriptsize
\begin{lstlisting}[basicstyle=\scriptsize,frame=single]
for (j = 0; j < N; j++) {
  for (i = 0; i <= M-2 - j; i++) {
    c[i][j] =  A[(M-2) - (i)][j]   / sqrt(A[(M-2) - (i)][j]*A[(M-2) - (i)][j]
			+ A[(M-2) - (i)+1][j]*A[(M-2) - (i)+1][j]);
    s[i][j] = -A[(M-2) - (i)+1][j] / sqrt(A[(M-2) - (i)][j]*A[(M-2) - (i)][j]
			+ A[(M-2) - (i)+1][j]*A[(M-2) - (i)+1][j]);
    for (k = j; k < N; k++) {
      A[(M-2) - (i)][k]   = c[i][j] * A[(M-2) - (i)][k] - s[i][j] * A[(M-2) - (i)+1][k];
      A[(M-2) - (i)+1][k] = s[i][j] * A[(M-2) - (i)][k] + c[i][j] * A[(M-2) - (i)+1][k];
    }
  }
}
\end{lstlisting}
}
\end{minipage}
\vspace{-.2cm}
\caption{\label{lst:givens-mod} Modified Givens Rotation before tiling}
\vspace{-.7cm}
\end{figure}

It
was obtained by first applying loop
normalization \cite{allenkennedybook} which consists in making all
loops iterate from $0$ to some greater value while appropriately
modifying the expressions involving the loop iterators within the loop
body. Then, we applied scalar expansion on \texttt{c}
and \texttt{s} \cite{allenkennedybook} to remove dependences induced
by those scalars which make loop permutation illegal. As a result, the
modified code is an affine code with fewer dependences, enabling it to
be automatically tiled by the Pluto compiler \cite{pluto}. The final
tiled code obtained, with default tile-sizes, can be found
in \cite{taco14TR}.

\paragraph{Performance comparison}

Fig.~\ref{fig:givens-comparison}-left shows the improvements in the
reuse distance profile using the convex partitioning heuristics and
the improved reuse distance profile of the tiled code. A better
profile is obtained for the tiled version than for convex
partitioning. Similarly to Matmult, this is because our convex
partitioning heuristic makes simplification for scalability and is
therefore not guaranteed to provide the best achievable reuse profile.

Fig~\ref{fig:givens-comparison}-right shows a two-fold improvement in
the performance of the transformed code for a matrix of size 4000.

\vspace{-.4cm}

\subsubsection{Householder Transformation}
\label{sec:cs-house}
\paragraph{Original program} Fig.~\ref{lst:householder} shows the 
original input code for the Householder transform, another approach
for QR decomposition.

\begin{figure}[h!tb]
\centering\begin{minipage}{.7\textwidth}
{\scriptsize
\begin{lstlisting}[basicstyle=\scriptsize,frame=single]
for (j = 0; j < N; j++) {
  total = 0;
  for (i = j+1; i < M; i++) {
    total += A[i][j] * A[i][j];
  }
  norm_x = (A[j][j] * A[j][j] + total);
  if (norm_x != 0) {
    if (A[j][j] < 0)
      norm_x = -norm_x;
    v[j] = norm_x + A[j][j];
    norm_v = (v[j] * v[j] + total);
    v[j] /= norm_v;
    for (i = j+1; i < M; i++) {
      v[i] = A[i][j] / norm_v;
    }
    for (jj = j; jj < N; jj++) {
      dot = 0.;
      for (kk = j; kk < M; kk++) {
	dot += v[kk] * A[kk][jj];
      }
      for (ii = j; ii < M; ii++) {
	A[ii][jj] -= 2 * v[ii] * dot;
      }
    }
  }
 }
\end{lstlisting}
}
\end{minipage}
\vspace{-.2cm}
\caption{\label{lst:householder}Householder computation}
\end{figure}
Comparing this with Givens (a different approach to compute the QR
decomposition) in terms of data locality potential is of interest: if
one has better locality potential than the other, then it would be
better suited for deployment on machines where the data movement cost
is the bottleneck. It complements complexity analysis, which only
characterizes the total number of arithmetic operations to be
performed. Indeed, on hardware where the computation power has become
increasingly cheaper relative to data access costs, standard
complexity analysis alone is insufficient to capture the relative
merits of alternative algorithms for a computation such as QR
decomposition.

\begin{figure}[h!tb]
\vspace{-.3cm}
  \centering
  \mbox{
    \subfloat{
      \begin{minipage}[b]{0.45\textwidth}
      \includegraphics[width=6.7cm]{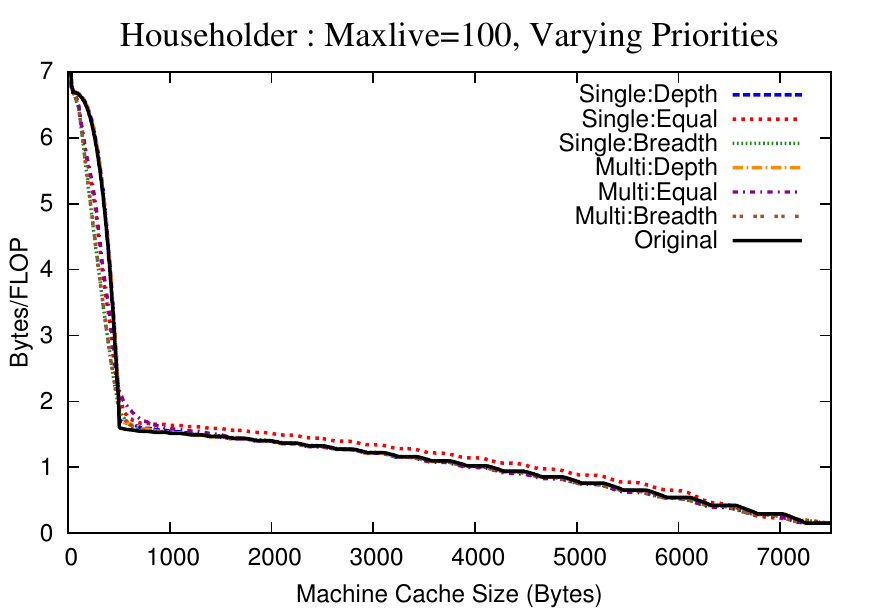}
      \label{fig:householder_priorities}
      \end{minipage}
    }
    \hspace{0.7cm}
   \subfloat{
      \begin{minipage}[b]{0.45\textwidth}
      \includegraphics[width=6.7cm]{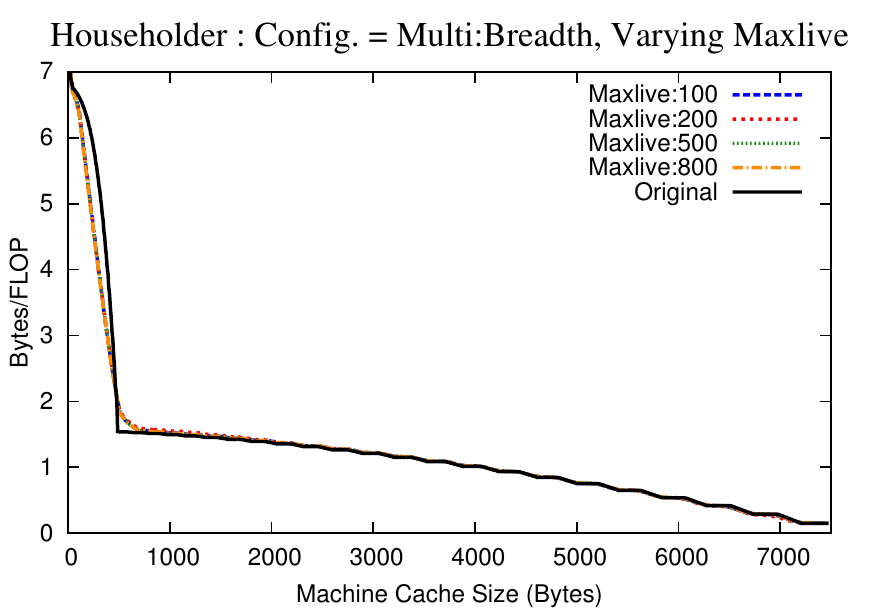}
      \label{fig:householder_cache_variation}
      \end{minipage}
    }
 }
\vspace{-.3cm}
\caption{Results with different heuristics for Householder}
\vspace{-.3cm}
\label{fig:householder}
\end{figure}

\paragraph{Analysis}
Fig.~\ref{fig:householder} shows the reuse distance profile of the
original code and the convex-partitioning for the Householder
algorithm, for an input matrix of size 30 by 30.

We observe a significant difference compared with Givens: the gap between
the reuse distance profile of the original code and that found by
our dynamic analysis is negligible. From this we conclude that the
potential for data locality improvement of this algorithm is limited,
and therefore we did not seek an optimized implementation for it.

Furthermore, comparing the bytes/flop required with the Givens graph
in Fig.~\ref{fig:givens-comparison}-left shows that our tiled
implementation of Givens achieves a significantly lower byte/flop
ratio, especially for small cache sizes. We conclude that the Givens
rotation algorithm may be better suited for deployment on future
hardware, because of its lower bandwidth demand than Householder,
especially for small cache sizes.

\subsubsection{Lattice-Boltzmann Method}
\label{sec:cs-lbm}
\paragraph{Original program} 


\emph{470.lbm}, a SPEC2006~\cite{SPEC2006,lbm} benchmark, 
implements the Lattice-Boltzmann technique to simulate fluid flow in 3
dimensions, in the presence of obstacles. The position and structure
of the obstacles are known only at run-time, but do not change
throughout the course of the computation.

\paragraph{Analysis} 

\begin{figure}[h!tb]
  \centering
  \mbox{ 
    \subfloat{
      \begin{minipage}[b]{0.45\textwidth}
      \includegraphics[width=6.7cm]{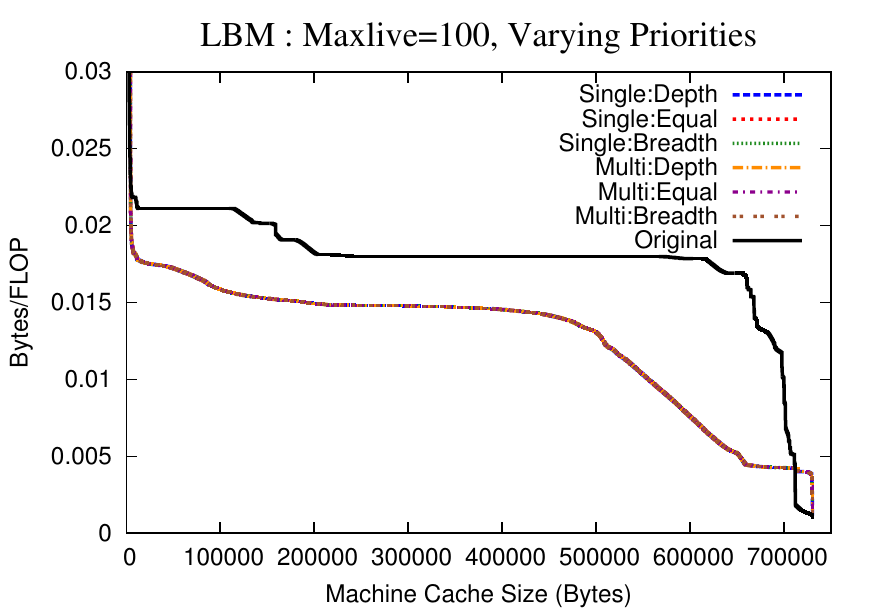}
      \label{fig:lbm_priorities}
      \end{minipage}
    }
    \hspace{0.7cm}
   \subfloat{
      \begin{minipage}[b]{0.45\textwidth}
      \includegraphics[width=6.7cm]{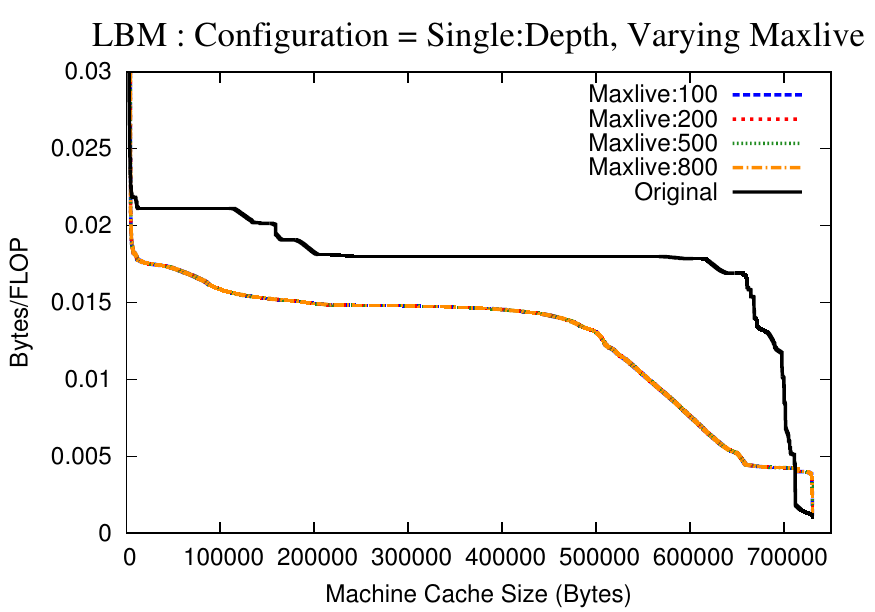}
      \label{fig:lbm_cache_variation}
      \end{minipage}
    }
 }
\vspace{-.5cm}
\caption{Results with different heuristics for 470.lbm}
\vspace{-.4cm}
\label{fig:lbm}
\end{figure}

The convex-partition heuristics account for the data dependent
behavior of the computation and are able to find valid operation
reordering with enhanced data locality, as shown
Fig.~\ref{fig:lbm}. The \emph{test} input size provided by the SPEC
benchmark suite was used for the analysis. To reduce the size of the
generated trace the problem size was reduced by a factor of $4$ along
each dimension. The reduced problem still has the same behavior as the
original problem.

For a cache size of $60$KB, the reordering after convex partitioning
obtained by the heuristics show an improvement in the Bytes/FLOP
ratio. For this benchmark all configurations of the heuristics yield
essentially identical results. However, unlike the previous
benchmarks, the absolute value of the bytes/flop is extremely low
(Fig.~\ref{fig:lbm}), indicating that the computation is already
compute-bound and a tiled version of the code would not be able to
achieve significant improvements in performance over the untiled
code. On an Intel Xeon E5640 with a clock speed of $2.53$GHz, the
untiled version already achieves a performance of $4$GFLOPS.  But
since the current trend in hardware architecture suggests that the
peak performance will continue to grow at a faster rate than the
increase in main-memory bandwidth, it is reasonable to expect that
optimizations like tiling that improve data locality will be critical
in the future even for such computations that are currently
compute-bound.

\vspace{-.3cm}

\subsubsection{410.bwaves}

\begin{figure}[h!tb]
  \centering
  \mbox{
    \subfloat{
      \begin{minipage}[b]{0.45\textwidth}
      \includegraphics[width=6.7cm]{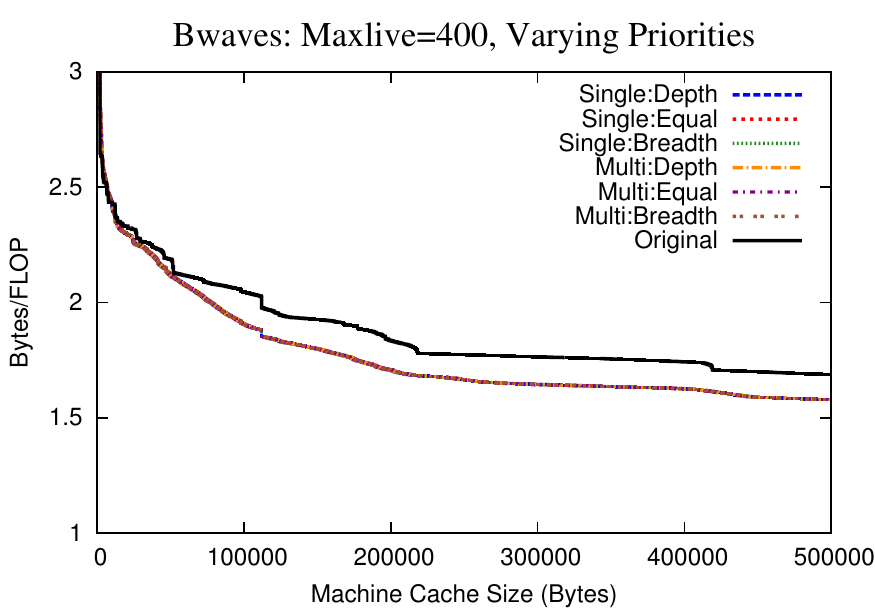}
      \label{fig:bwaves_priorities}
      \end{minipage}
    }
    \hspace{0.7cm}
   \subfloat{
      \begin{minipage}[b]{0.45\textwidth}
      \includegraphics[width=6.7cm]{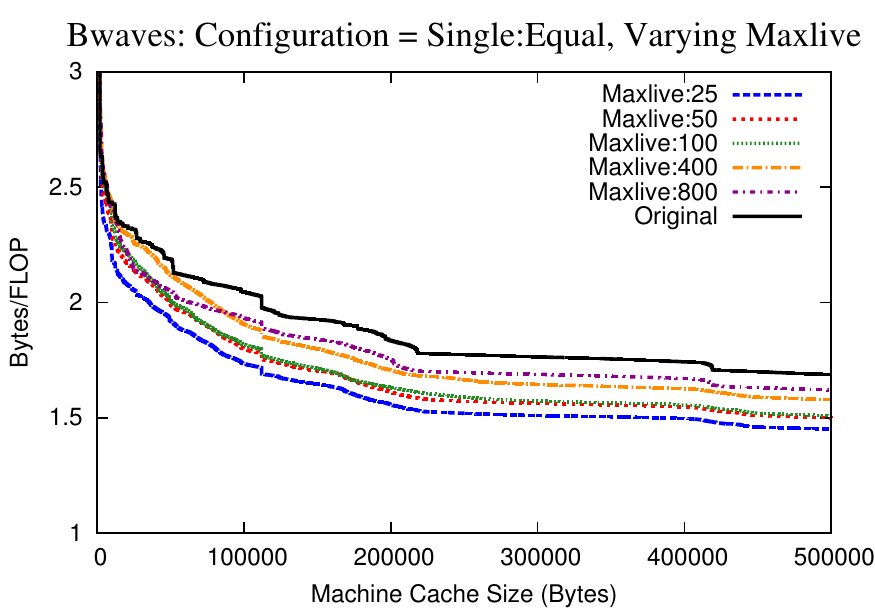}
      \label{fig:bwaves_cache_variation}
      \end{minipage}
    }
 }
\vspace{-.3cm}
\caption{Results with different heuristics for 410.bwaves}
\vspace{-.2cm}
\label{fig:bwaves}
\end{figure}

This benchmark is a computational fluid dynamic application from
SPEC2006~\cite{SPEC2006}.  We ran our analysis on the whole benchmark
with \emph{test} input size. For this benchmark too, the size of the
problem was reduced by a factor of $4$ along each dimension.  The
result of the analysis, shown in Fig.~\ref{fig:bwaves}, indicate a
limited potential for improving data locality.
\vspace{-.2cm}

\subsubsection{Large-Eddy Simulations with Linear-Eddy Model in 3D}

\begin{figure}[h!tb]
  \centering
  \mbox{
    \subfloat{
      \begin{minipage}[b]{0.45\textwidth}
      \includegraphics[width=6.7cm]{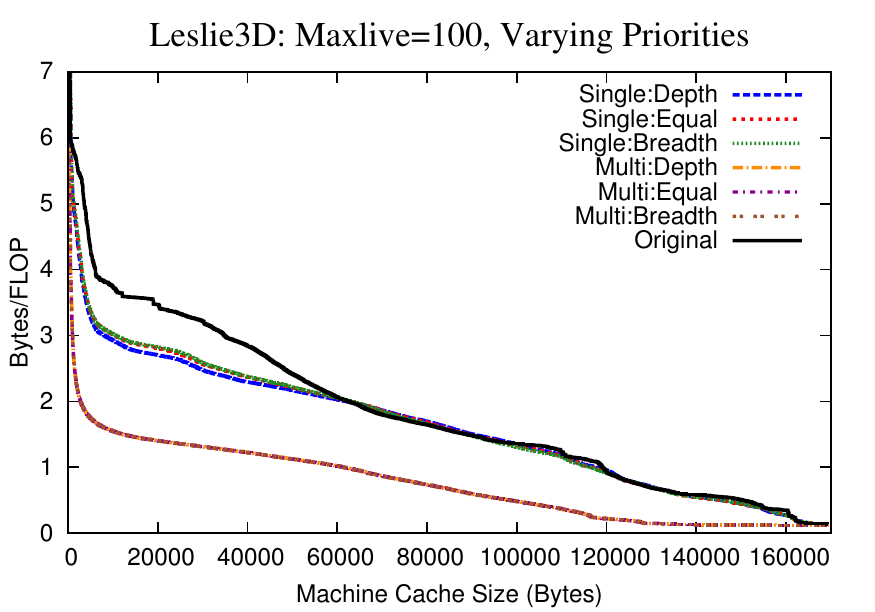}
      \label{fig:leslie_priorities}
      \end{minipage}
    }
    \hspace{0.7cm}
   \subfloat{
      \begin{minipage}[b]{0.45\textwidth}
      \includegraphics[width=6.7cm]{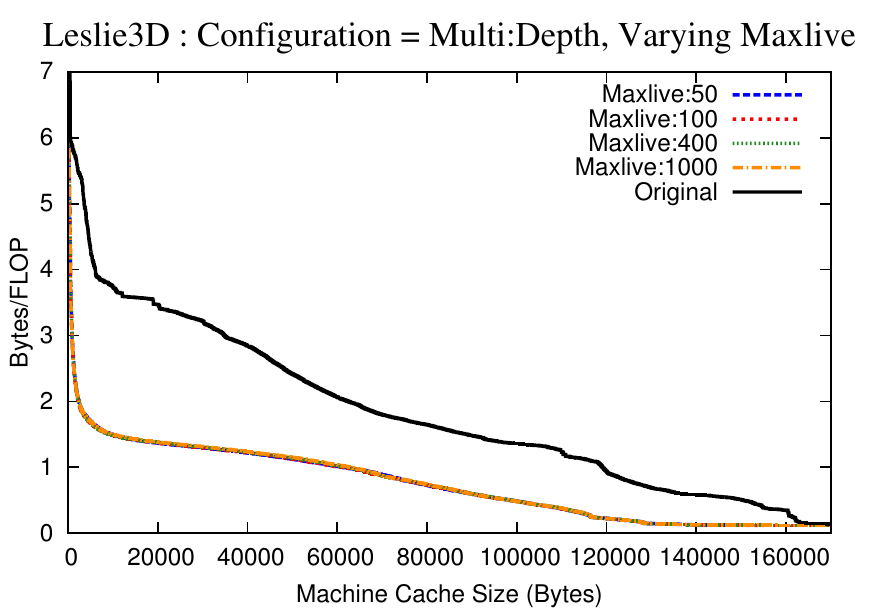}
      \label{fig:leslie_cache_variation}
      \end{minipage}
    }
 }
\vspace{-.3cm}
\caption{Results with different heuristics for 437.leslie3d}
\label{fig:leslie}
\end{figure}

437.leslie3d is another computational fluid dynamic benchmark from
SPEC2006~\cite{SPEC2006}. Here too, the analysis was done using
the \emph{test} dataset as given.
As shown in Fig.~\ref{fig:leslie}, leslie3d achieves a lower
bytes/FLOP ratio with the multi-level algorithm. The trend is not
sensitive to varying Maxlive.  Therefore, from the results, we
conclude that this benchmark has high potential for locality
improvement. We however leave for future work the task of deriving an
optimized implementation for leslie3d.

\newpage
\subsubsection{Odd-Even Sort}
\label{sec:odd-even}
\paragraph{Original program} Our dynamic analysis does not impose
any requirement on the data layout of the program: arrays, pointers,
structs etc. are seamlessly handled as the trace extraction tool
focuses exclusively on the address used in memory read/write
operations. To illustrate this we show in Fig.~\ref{lst:odd-even-orig}
the original code for an odd-even sorting algorithm, using a
linked-list implementation. \texttt{CompareSwap} compares the data
between two consecutive elements in the list, and swaps them if
necessary. Each swap operation is represented by a node in the CDAG.

\vspace{-.3cm}
\begin{figure}[h!tb]
\centering\begin{minipage}{0.8\textwidth}
{\scriptsize
\begin{lstlisting}[basicstyle=\scriptsize,frame=single]
for(i=0; i<N/2; ++i) {
  node *curr;
  for(curr=head->nxt; curr->nxt; curr=curr->nxt->nxt) {
    CompareSwap(curr, curr->nxt);
  }
  for(curr=head; curr; curr=curr->nxt->nxt) {
    CompareSwap(curr, curr->nxt);
  }
}
\end{lstlisting}
}
\end{minipage}
\vspace{-.3cm}
\caption{\label{lst:odd-even-orig} Odd-Even sort on linked list}
\vspace{-.5cm}
\end{figure}

\begin{figure}[h!tb]
\centering
\mbox{
	\subfloat{
		\begin{minipage}[b]{0.45\textwidth}
		\includegraphics[width=6.7cm]{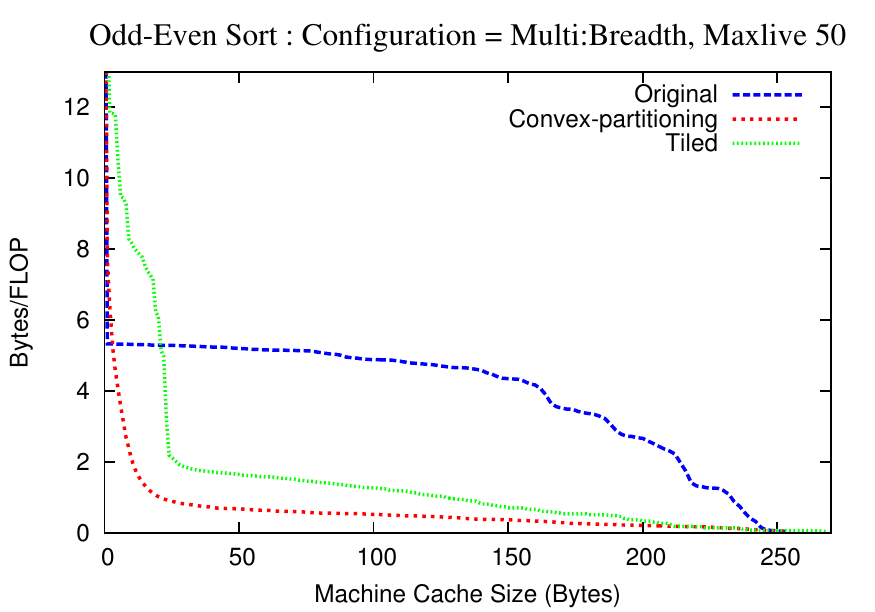}
			\end{minipage}
	}
	\hspace{0.5cm}
	\subfloat{
		\begin{minipage}[b]{0.45\textwidth}
		\includegraphics[width=6.7cm]{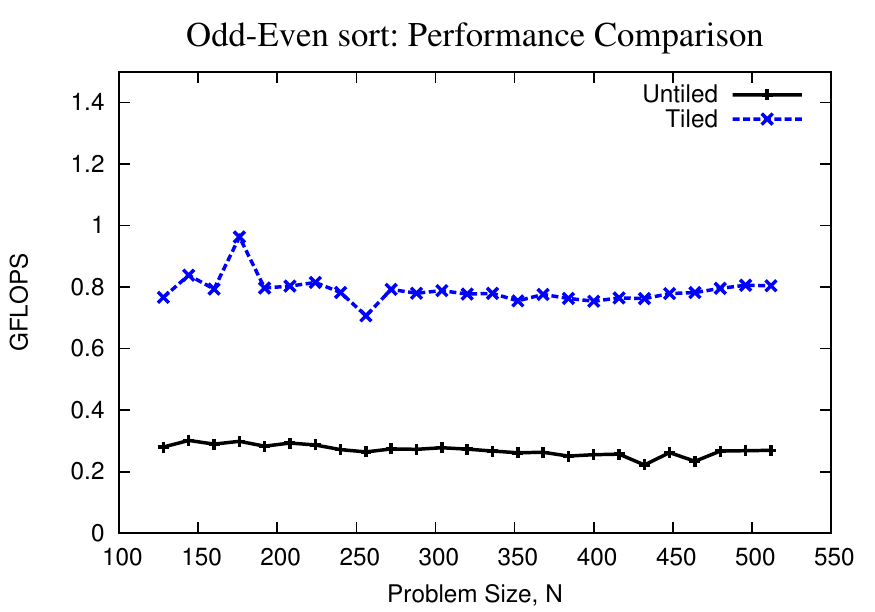}
			\end{minipage}
	}
}
\caption{\label{fig:odd-even-comparison}Odd-Even sort: Performance improvements due to tiling}
\end{figure}

\paragraph{Analysis}
We have performed our analysis on the original code, for a input list
of size 256, with random values. The profile of the original, best
convex partitioning (obtained with the best set of heuristic
parameters for this program) and tiled (our modified) implementation
are shown shown in the left plot in Fig.~\ref{fig:odd-even-comparison}. More
complete analysis of the various heuristic parameters result can be
found in \cite{taco14TR}.



\paragraph{Modified implementation}
Based on careful analysis of the original code in
Fig.~\ref{lst:odd-even-orig}, an equivalent register-tiled version
with a tile size of 4 was manually developed. It can be found in \cite{taco14TR}.

\paragraph{Performance comparison}

The comparison of performance of the untiled and tiled versions of the
code is shown in Fig.~\ref{fig:odd-even-comparison}.  The left plot
shows the improved data locality for the tiled code compared to the
original code. The actual improvement in performance of the tiled code
is shown in the right plot in Fig.~\ref{fig:odd-even-comparison} for a
random input. Experiments about sensitivity to datasets reported in
later Sec.~\ref{subsec:datasets} confirm that our optimized variant
consistently outperform the original code.

\subsubsection{LU Decomposition (LAPACK)}
\label{sec:lud}

The last benchmark we analyze is an implementation of the LU
decomposition for dense matrices, from the LAPACK
package \cite{lapackbook}. It uses pivoting (therefore the computation
is input-dependent) and LAPACK provides both a base implementation
meant for small problem sizes, and a block decomposition for large
problem sizes \cite{lapackweb}. Both code versions can be found
in \cite{lapackbook}.



We have run our dynamic analysis on the non-blocked implementation of
LU decomposition, for a single random matrix of size 128 by 128. The results of varying heuristics and maxlive can be found in \cite{taco14TR}.

\vspace{-.3cm}
\begin{figure}[h!tb]
  \centering
  \mbox{
    \subfloat{
      \begin{minipage}[b]{0.45\textwidth}
      \includegraphics[width=6.7cm]{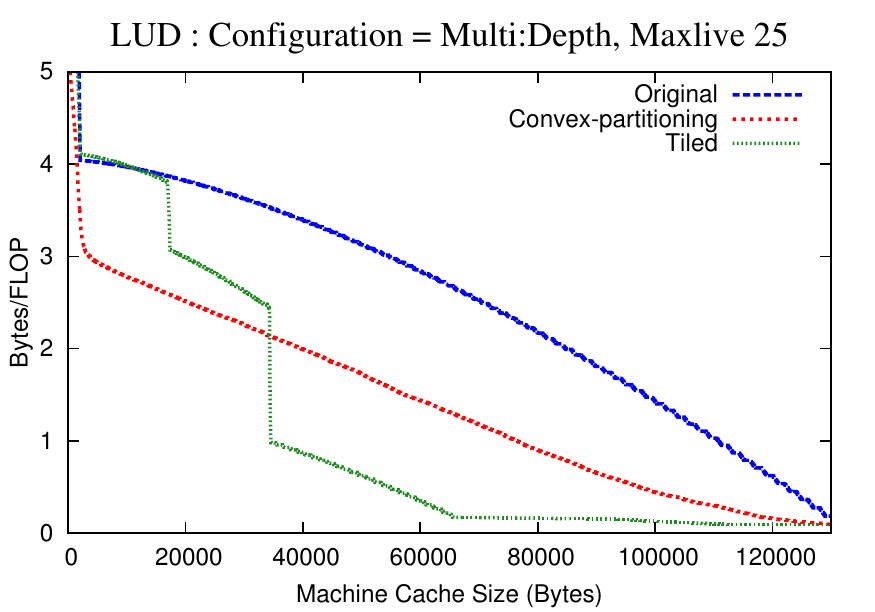}
      \end{minipage}
    }
    \hspace{0.7cm}
   \subfloat{
      \begin{minipage}[b]{0.45\textwidth}
      \includegraphics[width=6.7cm]{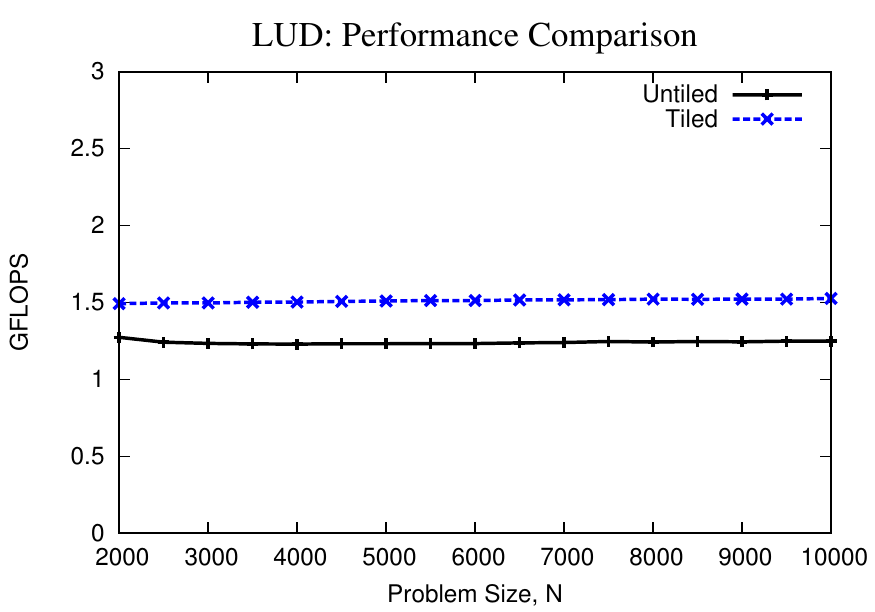}
      \end{minipage}
    }
 }
\vspace{-.15cm}
\caption{LU Decomposition}
\vspace{-.3cm}
\label{fig:lud1}
\end{figure}

%
%
A potential for locality improvement is established from the result of
the analysis. On Fig.~\ref{fig:lud1}-left, we plot the reuse distance
profile of the non-blocked implementation; the convex partitioning
with the best heuristic parameters for this program; and tiled (i.e.,
blocked) implementation, in addition to the original (i.e.,
non-blocked) one. The blocked version shows a better bytes/Flop than
the convex partitioning for cache sizes larger than 35kB. This is
likely due to inter-block reuse achieved in the highly optimized
LAPACK implementation, combined with a sub-optimal schedule found by
our heuristic. Further, Fig.~\ref{fig:lud1}-right shows actual
performance, in GFLOPS, for non-blocked and blocked versions of the
code.

%
%

\subsection{Dataset Sensitivity Experiments}
\label{subsec:datasets}


We conclude our experimental analysis with a study of the sensitivity
of our analysis to different datasets. In the set of benchmarks
presented in the previous section, the majority of them have a CDAG
that depends only on the input dataset size, and not on the input
values. Therefore for these codes our dynamic analysis results hold
for any dataset of identical size.

\begin{figure}[h!tb]
\centering
\begin{minipage}{.99\textwidth}
\begin{minipage}{.48\textwidth}
\includegraphics[width=6.7cm]{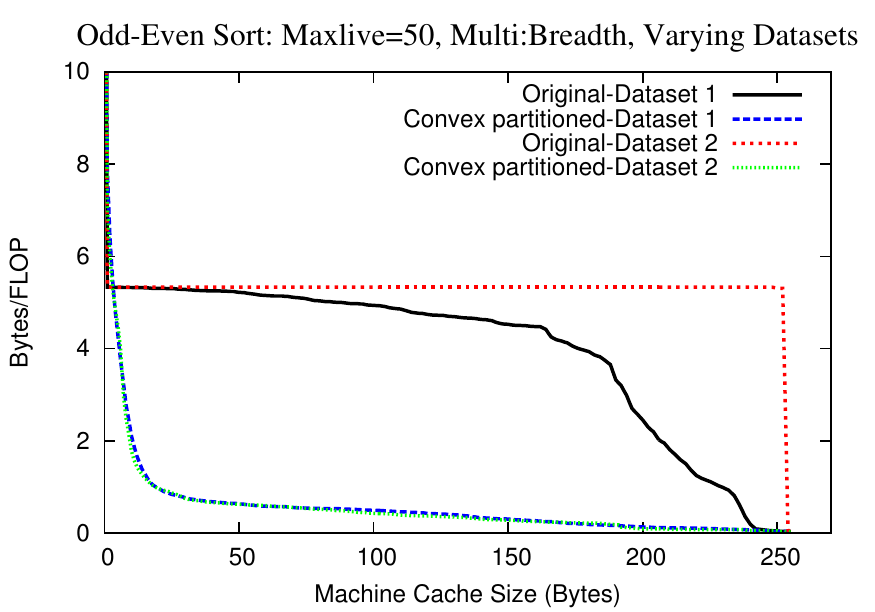}
\caption{\label{fig:odd-even-diff-inputs}Sensitivity analysis for
	odd-even sort}
\end{minipage}
\begin{minipage}{.48\textwidth}
\includegraphics[width=6.7cm]{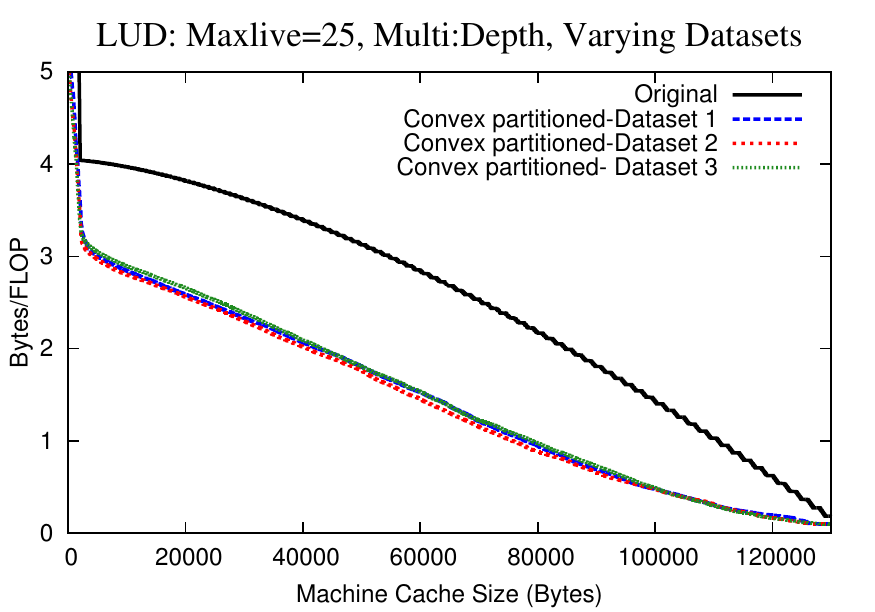}
\caption{\label{fig:lud-datasets}Sensitivity analysis for LUD}
\end{minipage}
\end{minipage}
\end{figure}

 Odd-even sort and LUD are two benchmarks that are
input-dependent. To determine the sensitivity of the
convex-partitioning heuristics on the input, we ran the heuristics on
different datasets, as shown in Fig.~\ref{fig:odd-even-diff-inputs}
and Fig.~\ref{fig:lud-datasets}.

Fig.~\ref{fig:odd-even-diff-inputs} shows the result for two such
datasets - one with random input elements and the other with the
reverse sorted input list, a worst-case dataset. We used multi-level
heuristics with breadth-first priority, which corresponds to the
parameters of the best result, which can be found in \cite{taco14TR}.
Fig.~\ref{fig:odd-even-diff-inputs} shows that the potential for improvement
exhibited by the heuristics remains consistent with varying inputs, i.e., 
the ``Convex partitioned'' reuse distance profile does not vary with the input
value. We note that in the general case, some variations are expected
for different datasets. Similarl to complexity analysis of such
algorithms, one needs to perform both worst-case analysis (i.e.,
reverse-sorted) and analysis on random/representative datasets for
best results.

Fig.~\ref{fig:lud-datasets} exhibits a similar behavior where the
three tested datasets have a similar analysis profile. The first
dataset has is a random matrix, the second was created so that the
pivot changes for about half the rows of the matrix, and for the third
one the pivot changes for all rows of the matrix. 

\begin{figure}[h!tb]
\centering
\begin{minipage}{.99\textwidth}
\begin{minipage}{.48\textwidth}
\includegraphics[width=6.7cm]{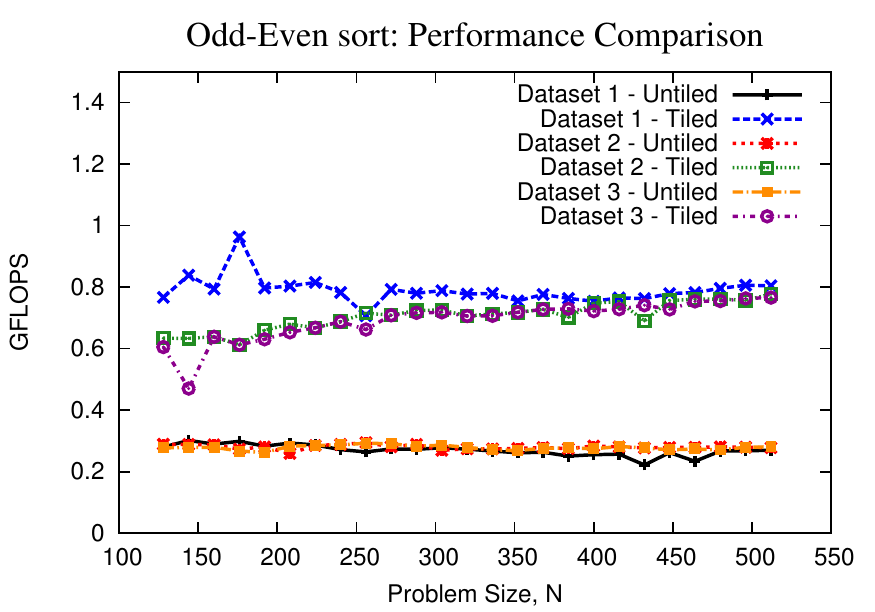}
\vspace{-.4cm}
\caption{\label{fig:oe-perf1}Performance for
	odd-even sort}
\end{minipage}
\begin{minipage}{.48\textwidth}
\includegraphics[width=6.7cm]{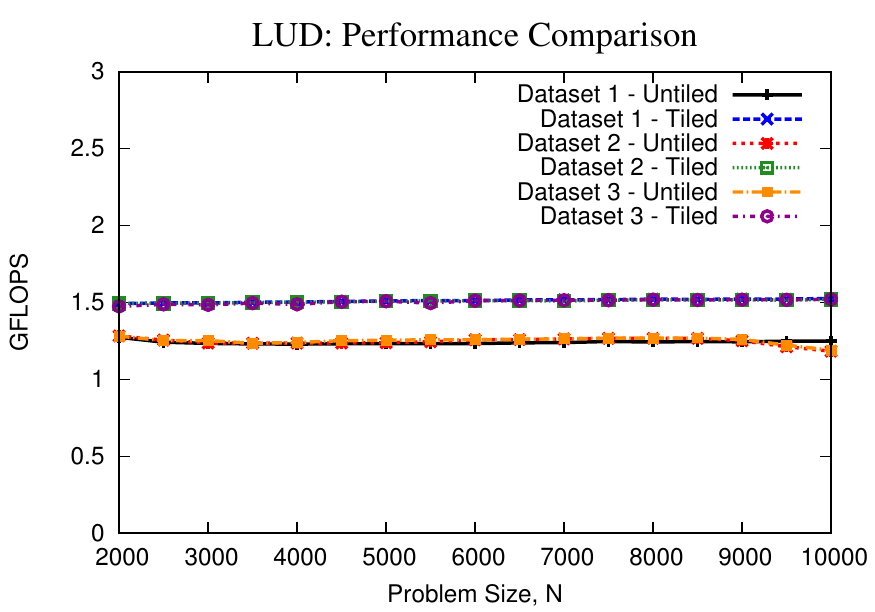}
\vspace{-.4cm}
\caption{\label{fig:lud-perf1}Performance for LUD}
\end{minipage}
\end{minipage}
\end{figure}

Finally, we complete our analysis by reporting in
Fig.~\ref{fig:oe-perf1}-\ref{fig:lud-perf1} the performance of our
manually optimized programs for odd-even sorting and LU
decomposition. For two programs (\textsf{Original}, and our modified
implementation \textsf{Tiled}), we plot the performance for various
datasets (different curves) and various sizes for those datasets
(different points on the x axis). We observe very similar asymptotic
performance for the various datasets.

\vspace{-.3cm}
\vspace{-.5cm}

\section{Related Work}
\label{sec:related}

Both algorithmic approaches
(e.g., \cite{demmel.unpub.08,BDHS11a,BDHS11b}) and compiler
transformations (e.g.,
\cite{irigoin88popl,wolf.91.pldi,kennedy.93.lcpc,uday08pldi}) have
been employed to improve data locality. The applicability of these
techniques to arbitrary computations is limited.  For example,
compiler optimizations typically require programs in which precise static
characterization of the run-time behavior is possible; this is 
challenging in the presence of inter-procedural control flow, diverse
data structures, aliasing, etc. Reuse distance
analysis~\cite{mattson1970valuation,ding2003predicting}, which
considers the actual observed run-time behavior, is more generally
applicable and has been used for cache miss rate
prediction~\cite{Zhong:pact03,Marin:sigmetrics2004,JiangZTS10},
program phase detection~\cite{Shen:asplos2004}, data layout
optimizations~\cite{Zhong:pldi2004}, virtual memory
management~\cite{Cascaval:pact2005}, and I/O performance
optimizations~\cite{Jiang:TC2005}.  Even though reuse distance
analysis provides insights into the data locality of software
behavior, it has been limited to analyzing locality characteristics
for a specific execution order of the operations, typically that
generated by a sequential program. 
In contrast to all previous
work on reuse distance analysis, to the best of our knowledge, our
upper-bounding approach is the first to attempt a schedule-independent
characterization of the inherent data locality characteristics of
a CDAG.


The idea of considering valid re-orderings of a given execution trace
has been applied successfully to characterize the potential
parallelism of applications.  Kumar's approach \cite{kumar-toc88} 
computes a possible valid schedule using a timestamping analysis of
an instrumented statement-level execution of the sequential program.
Shadow variables are used to store the last modification times for
each variable.  Each run-time instance of a statement is associated
with a timestamp that is one greater than the last-modify times of all
its operands.  A histogram of the number of
operations at each time value provides a fine-grained parallelism
profile of the computation, and the maximal timestamp represents the
critical path length for the entire computation. Other prior efforts
with a similar overall approach include
\cite{austin-isca92,nicolau-toc84,kumar-toc88,wall-asplos91,lam-isca92,theobald-micro92,rauchwerger-micro93,postiff-sigarch99,stefanovic-europar00,mak-woda09,Garcia-PLDI2011}.


In contrast to the above fine-grained approach, an alternate technique
developed by Larus \cite{larus-tpds93} performed analysis of
loop-level parallelism at different levels of a nested loop. Loop-level
parallelism is measured by forcing a sequential order of execution of
statements within each iteration of a loop being characterized, so
that the only available concurrency is across different iterations of
that loop. A related technique is applied in the context of speculative
parallelization of loops, where dynamic dependences across loop
iterations are tracked \cite{rauchwerger-pldi95}. 
A few recent
approaches of similar nature include
\cite{bridges-micro07,tian-micro08,zong-hpca08,wu-lcpc08,oancea-lcpc08,tournavitis-pldi09}.
In order to estimate parallel speedup of DAGs, Sarkar and Hennessy
\cite{sarkar86b} developed convex partitioning of DAGs.  In previous
work, we
\cite{holewinski-pldi2012} used dynamic analysis of CDAGs
to assess the vectorization potential of codes that are not effectively
vectorized by current vectorizing compilers. However, we are not aware of
any prior work on dynamic analysis of CDAGs with the goal of
characterizing and/or enhancing data locality properties of computations.

\section{Discussion}
\label{sec:discussion}

In this section, we discuss the potential and some of the current
limitations of the dynamic analysis approach for data locality
characterization/enhancement that we have developed in this article.

\paragraph{Dependence on Input Values}

As with any work that uses dynamic analysis of the actual execution
trace of a program, any conclusions drawn are only strictly true for
that particular execution. For programs where the execution trace is
dependent on input data, the CDAG will change for different runs. 
Due to space limitations, we only
present RDA results for a single problem size for each benchmark. However,
we have experimented with different problem sizes and the qualitative
conclusions remain stable across problem size. Further, as demonstrated
by the case studies of the Floyd-Warshall and Givens rotation codes,
the modified codes based on insights from the dynamic analysis were
demonstrated to exhibit consistent performance improvement for different
problem sizes.

\paragraph{Overhead of Analysis} 

The initial prototype implementation of the partitioning algorithm has
not yet been optimized and currently has a fairly high overhead (about
4 orders of magnitude) compare to the execution time. As discussed
earlier, the computational complexity of the partitioning algorithm is
$O(|T|)$ for the single-level version, and $O(|T| \log(|M|))$ for the
multi-level version and can therefore be made sufficiently efficient
to be able to analyze large-scale applications in their entirety.

\paragraph{Trace Size Limitations}

A more significant challenge and limitation of the current system we
implemented is the memory requirement.  The CDAG size is usually a
function of the problem size. For instance, for 470.lbm using the
\emph{test} dataset as is would generate a CDAG of size 120GB. The
additional data structures used by the tool quickly exhaust the memory
on our machines.  For instance for 470.lbm we were able to easily
create a smaller input dataset (e.g., 1/64th) leading to a 3GB CDAG,
that our tool could handle. However, there are numerous benchmarks
where such reduction of the input dataset is not possible, and/or does
not affect the CDAG size. For those codes, where the CDAG was beyond a
few GBs, our current implementation fails due to insufficient memory.
The development of an ``out-of-core'' analyzer
is the focus of ongoing follow-up work. Another solution is to
compress the CDAG using a technique similar to trace
compression~\cite{ketterlin-cgo2008} leading to a space complexity of
$O(|M|+|P|)$ in the most favorable scenario (in which all dependences
turn out to be affine). More generally trace sampling techniques can
be applied to tackle scalability issues of this approach.


\paragraph{Tightness of Estimation} 

The primary goal of our CDAG partitioning algorithm is to find a more
favorable valid reordering of the schedule of its operations so as to
lower the volume of data movement between main memory and cache. From
the perspective of data movement, the lowest possible amount, which
corresponds to the best possible execution order for a given CDAG and
a given cache size, can be considered the {\em inherent data access
  complexity} of that computation.  Irrespective of how much lower the
reordered schedule's data movement volume is compared to the original
schedule's data movement volume, how do we determine how close we are
to the best possible valid order of execution? A possible solution is
to work from the opposite direction and develop lower bound techniques
for the data access complexity of CDAGs. In a complementary work we
are developing an approach to establishing lower bounds on the data
movement costs for arbitrary CDAGs. One way of assessing the tightness
of the upper bounds (this work) and lower bounds (complementary work
in progress) is to see how close these two bounds are for a given
CDAG.

\paragraph{Use of Analysis} 

We envision several uses of a tool based on the dynamic analysis
approach developed in this paper. (1)
 {\em For Application Developers}: By running the dynamic analysis on
  different phases of an application, along with standard performance
  profiling, it is possible to identify which of the computationally
  dominant phases of the application may have the best potential for
  performance improvement through code changes that enhance data
  reuse. Dynamic analysis for data locality can also be used to
  compare and choose between alternate equivalent algorithms with the
  same functionality -- even if they have similar performance on
  current machines. If the reuse distance profiles of two algorithms
  after reordering based on dynamic analysis are very different, the
  algorithm with lower bandwidth demands would likely be better for
  the future.
(2) {\em For Compiler Developers}: Compilers implement many
  transformations like fusion and tiling that enhance data reuse. The
  results from the convex partitioning of the CDAG can be used to
  gauge the impact of compiler transformations and potential for improvement.
(3) {\em For Architecture Designers}: Running the dynamic analysis tool on
  a collection of representative applications can guide vendors of
  architectures in designing hardware that provides adequate bandwidth
  and/or sufficient capacity for the different levels of the memory
  hierarchy.

\section{Conclusion}
\label{sec:conc}

With future systems, the cost of data movement through the memory
hierarchy is expected to become even more dominant relative to the
cost of performing arithmetic operations
\cite{bergman2008exascale,fuller2011,shalf2011exascale}, both in terms
of throughput and energy. Therefore optimizing data locality will
become ever more critical in the coming years. Given the crucial
importance of optimizing data access costs in systems with
hierarchical memory, it is of great interest to develop tools and
techniques for characterization and enhancement of the data locality
properties of an algorithm.  Although reuse distance
analysis~\cite{mattson1970valuation,ding2003predicting} provides a
useful characterization of data locality for a given execution trace,
it fails to provide any information on the potential for improving the
in data reuse through valid reordering of the operations in the
execution trace.

In this paper, we have developed a dynamic analysis approach to
provide insights beyond what is possible from standard reuse distance
analysis.  Given an execution trace from a sequential program, we seek
to (i) characterize the data locality properties of an algorithm and
(ii) determine if there exists potential for enhancement of data
locality through execution reordering.  Since we first explicitly
construct a dynamic computational directed acyclic graph (CDAG) to
capture the statement instances and their inter-dependences; perform
convex partitioning of the CDAG to generate a modified,
dependence-preserving, execution order with better expected data
reuse; and then perform reuse distance analysis on the trace
corresponding to the modified execution order, we expect to get a
better characterization of the potential benefit of reordering. We
have demonstrated the utility of the approach in
characterizing/enhancing data locality for a number of benchmarks.

\bibliographystyle{plain}
\bibliography{mypaper}

\begin{thebibliography}{10}

\bibitem{lapackbook}
E.~Anderson, Z.~Bai, C.~Bischof, S.~Blackford, J.~Demmel, J.~Dongarra,
  J.~Du~Croz, A.~Greenbaum, S.~Hammarling, A.~McKenney, and D.~Sorensen.
\newblock {\em {LAPACK} Users' Guide}.
\newblock Society for Industrial and Applied Mathematics, Philadelphia, PA,
  third edition, 1999.

\bibitem{austin-isca92}
Todd Austin and Gurindar Sohi.
\newblock Dynamic dependency analysis of ordinary programs.
\newblock In {\em ISCA}, pages 342--351, 1992.

\bibitem{BDHS11b}
G.~Ballard, J.~Demmel, O.~Holtz, and O.~Schwartz.
\newblock Minimizing communication in numerical linear algebra.
\newblock {\em SIAM Journal on Matrix Analysis and Applications},
  32(3):866--901, 2011.

\bibitem{BDHS11a}
Grey Ballard, James Demmel, Olga Holtz, and Oded Schwartz.
\newblock Graph expansion and communication costs of fast matrix
  multiplication: regular submission.
\newblock In {\em Proc. {SPAA}}, pages 1--12, New York, NY, USA, 2011. ACM.

\bibitem{bergman2008exascale}
K.~Bergman, S.~Borkar, et~al.
\newblock Exascale computing study: {T}echnology challenges in achieving
  exascale systems.
\newblock {\em {DARPA IPTO}, {T}ech. {R}ep}, 2008.

\bibitem{bilardi2001characterization}
G.~Bilardi and E.~Peserico.
\newblock A characterization of temporal locality and its portability across
  memory hierarchies.
\newblock {\em Proc. {ICALP}}, pages 128--139, 2001.

\bibitem{uday08pldi}
Uday Bondhugula, Albert Hartono, J.~Ramanujan, and P.~Sadayappan.
\newblock A practical automatic polyhedral parallelizer and locality optimizer.
\newblock In {\em Proc. {PLDI}}, 2008.

\bibitem{bridges-micro07}
Matthew Bridges, Neil Vachharajani, Yun Zhang, Thomas Jablin, and David August.
\newblock Revisiting the sequential programming model for multi-core.
\newblock In {\em MICRO}, pages 69--84, 2007.

\bibitem{Cascaval:pact2005}
Calin Cascaval, Evelyn Duesterwald, Peter~F. Sweeney, and Robert~W. Wisniewski.
\newblock Multiple page size modeling and optimization.
\newblock In {\em Proc. {PACT}}. IEEE Computer Society, 2005.

\bibitem{demmel.unpub.08}
J.~Demmel, L.~Grigori, M.~Hoemmen, and J.~Langou.
\newblock Communication-optimal parallel and sequential {QR} and {LU}
  factorizations.
\newblock {\em SIAM Journal on Scientific Computing}, 34(1):206--239, 2012.

\bibitem{ding2003predicting}
Chen Ding and Yutao Zhong.
\newblock {Predicting whole-program locality through reuse distance analysis}.
\newblock In {\em PLDI}, pages 245--257. ACM, 2003.

\bibitem{taco14TR}
Naznin Fauzia, Venmugil Elango, Mahesh Ravishankar, Louis-No{\"e}l Pouchet,
  J.~Ramanujam, Fabrice Rastello, Atanas Rountev, and P.~Sadayappan.
\newblock Beyond reuse distance analysis: Dynamic analysis for characterization
  of data locality potential.
\newblock Technical Report OSU-CISRC-9/13-TR19, Ohio State University,
  September 2013.

\bibitem{fuller2011}
Samuel~H. Fuller and Lynette~I. Millett.
\newblock {\em The Future of Computing Performance: Game Over or Next Level?}
\newblock The National Academies Press, 2011.

\bibitem{Garcia-PLDI2011}
Saturnino Garcia, Donghwan Jeon, Christopher~M. Louie, and Michael~Bedford
  Taylor.
\newblock Kremlin: {Rethinking} and rebooting gprof for the multicore age.
\newblock In {\em PLDI}, pages 458--469, 2011.

\bibitem{Hennessey-Patterson2011}
J.L. Hennessy and D.A. Patterson.
\newblock {\em Computer architecture: a quantitative approach}.
\newblock Morgan Kaufmann, 2011.

\bibitem{SPEC2006}
John~L. Henning.
\newblock Spec cpu2006 benchmark descriptions.
\newblock {\em SIGARCH Comput. Archit. News}, 2006.

\bibitem{holewinski-pldi2012}
Justin Holewinski, Ragavendar Ramamurthi, Mahesh Ravishankar, Naznin Fauzia,
  Louis-No\"{e}l Pouchet, Atanas Rountev, and P.~Sadayappan.
\newblock Dynamic trace-based analysis of vectorization potential of
  applications.
\newblock In {\em Proc. {PLDI}}, pages 371--382, New York, NY, USA, 2012. ACM.

\bibitem{irigoin88popl}
F.~Irigoin and R.~Triolet.
\newblock Supernode partitioning.
\newblock In {\em Proc. {POPL}}, pages 319--329, 1988.

\bibitem{Jiang:TC2005}
Song Jiang and Xiaodong Zhang.
\newblock {Making LRU Friendly to Weak Locality Workloads: A Novel Replacement
  Algorithm to Improve Buffer Cache Performance}.
\newblock {\em IEEE Trans. Comput.}, 54:939--952, August 2005.

\bibitem{JiangZTS10}
Yunlian Jiang, Eddy~Z. Zhang, Kai Tian, and Xipeng Shen.
\newblock Is reuse distance applicable to data locality analysis on chip
  multiprocessors?
\newblock In {\em Proc. {Comp. Const.}}, pages 264--282, 2010.

\bibitem{allenkennedybook}
K.~Kennedy and J.~Allen.
\newblock {\em Optimizing compilers for modern architectures: A
  dependence-based approach}.
\newblock Morgan Kaufmann, 2002.

\bibitem{kennedy.93.lcpc}
Ken Kennedy and Kathryn~S. McKinley.
\newblock Maximizing loop parallelism and improving data locality via loop
  fusion and distribution.
\newblock In {\em Languages and Compilers for Parallel Computing}, pages
  301--320. Springer-Verlag, 1993.

\bibitem{ketterlin-cgo2008}
Alain Ketterlin and Philippe Clauss.
\newblock Prediction and trace compression of data access addresses through
  nested loop recognition.
\newblock In {\em Proc. {CGO}}, pages 94--103, 2008.

\bibitem{kumar-toc88}
Manoj Kumar.
\newblock Measuring parallelism in computation-intensive scientific/engineering
  applications.
\newblock {\em IEEE Transactions on Computers}, 37(9):1088--1098, September
  1988.

\bibitem{lam-isca92}
M.~Lam and R.~Wilson.
\newblock Limits of control flow on parallelism.
\newblock In {\em ISCA}, pages 46--57, 1992.

\bibitem{lapackweb}
{LAPACK}.
\newblock http://www.netlib.org/lapack.

\bibitem{larus-tpds93}
James Larus.
\newblock Loop-level parallelism in numeric and symbolic programs.
\newblock {\em IEEE Transactions on Parallel and Distributed Systems},
  4(1):812--826, July 1993.

\bibitem{mak-woda09}
Jonathan Mak and Alan Mycroft.
\newblock Limits of parallelism using dynamic dependency graphs.
\newblock In {\em WODA}, pages 42--48, 2009.

\bibitem{Marin:sigmetrics2004}
Gabriel Marin and John Mellor-Crummey.
\newblock Cross-architecture performance predictions for scientific
  applications using parameterized models.
\newblock In {\em SIGMETRICS '04/Performance '04}. ACM, 2004.

\bibitem{mattson1970valuation}
R.L. Mattson, J.~Gecsei, D.~Slutz, and I.~L Traiger.
\newblock {Evaluation techniques for storage hierarchies}.
\newblock {\em IBM Systems Journal}, 9(2):78--117, 1970.

\bibitem{nicolau-toc84}
A.~Nicolau and J.~Fisher.
\newblock Measuring the parallelism available for very long instruction word
  architectures.
\newblock {\em IEEE Transactions on Computers}, 33(11):968--976, 1984.

\bibitem{parda}
Qingpeng Niu, J.~Dinan, Qingda Lu, and P.~Sadayappan.
\newblock Parda: A fast parallel reuse distance analysis algorithm.
\newblock In {\em Proc. {IPDPS}}, pages 1284 --1294, may 2012.

\bibitem{oancea-lcpc08}
Cosmin Oancea and Alan Mycroft.
\newblock Set-congruence dynamic analysis for thread-level speculation ({TLS}).
\newblock In {\em LCPC}, pages 156--171, 2008.

\bibitem{prasanna-FW}
Joon-Sang Park, Michael Penner, and Viktor~K. Prasanna.
\newblock {Optimizing Graph Algorithms for Improved Cache Performance}.
\newblock {\em IEEE Transactions on Parallel Distributed Systems},
  15(9):769--782, 2004.

\bibitem{pluto}
{{Pluto}: A polyhedral automatic parallelizer and locality optimizer for
  multicores}.
\newblock http://pluto-compiler.sourceforge.net.

\bibitem{lbm}
Thomas Pohl.
\newblock $470.$lbm.
\newblock http://www.spec.org/cpu2006/Docs/470.lbm.html.

\bibitem{postiff-sigarch99}
Matthew Postiff, David Greene, Gary Tyson, and Trevor Mudge.
\newblock The limits of instruction level parallelism in {SPEC95} applications.
\newblock {\em SIGARCH Computer Architecture News}, 27(1):31--34, 1999.

\bibitem{rauchwerger-micro93}
Lawrence Rauchwerger, Pradeep Dubey, and Ravi Nair.
\newblock Measuring limits of parallelism and characterizing its vulnerability
  to resource constraints.
\newblock In {\em MICRO}, pages 105--117, 1993.

\bibitem{rauchwerger-pldi95}
Lawrence Rauchwerger and David Padua.
\newblock The {LRPD} test: {S}peculative run-time parallelization of loops with
  privatization and reduction parallelization.
\newblock In {\em PLDI}, pages 218--232, 1995.

\bibitem{sarkar86b}
Vivek Sarkar and John~L. Hennessy.
\newblock Compile-time partitioning and scheduling of parallel programs.
\newblock In {\em SIGPLAN Symposium on Compiler Construction}, pages 17--26,
  1986.

\bibitem{shalf2011exascale}
J.~Shalf, S.~Dosanjh, and J.~Morrison.
\newblock Exascale computing technology challenges.
\newblock {\em High Performance Computing for Computational Science--VECPAR
  2010}, pages 1--25, 2011.

\bibitem{Shen:asplos2004}
Xipeng Shen, Yutao Zhong, and Chen Ding.
\newblock Locality phase prediction.
\newblock In {\em Proc. {ASPLOS}}. ACM, 2004.

\bibitem{stefanovic-europar00}
Darko Stefanovi{\'{c}} and Margaret Martonosi.
\newblock Limits and graph structure of available instruction-level
  parallelism.
\newblock In {\em Euro-Par}, pages 1018--1022, 2000.

\bibitem{theobald-micro92}
Kevin Theobald, Guang Gao, and Laurie Hendren.
\newblock On the limits of program parallelism and its smoothability.
\newblock In {\em MICRO}, pages 10--19, 1992.

\bibitem{tian-micro08}
Chen Tian, Min Feng, Vijay Nagarajan, and Rajiv Gupta.
\newblock Copy or discard execution model for speculative parallelization on
  multicores.
\newblock In {\em MICRO}, pages 330--341, 2008.

\bibitem{tournavitis-pldi09}
Georgios Tournavitis, Zheng Wang, Zheng, Bj{\"{o}}rn Franke, and Michael
  O'Boyle.
\newblock Towards a holistic approach to auto-parallelization.
\newblock In {\em PLDI}, pages 177--187, 2009.

\bibitem{sahni-FW}
G.~Venkataraman, S.~Sahni, and S.~Mukhopadhyaya.
\newblock {A Blocked All-Pairs Shortest-Paths Algorithm}.
\newblock {\em Journal of Experimental Algorithmics}, 8:2.2, December 2003.

\bibitem{wall-asplos91}
David Wall.
\newblock Limits of instruction-level parallelism.
\newblock In {\em ASPLOS}, pages 176--188, 1991.

\bibitem{wolf.91.pldi}
Michael~E. Wolf and Monica~S. Lam.
\newblock A data locality optimizing algorithm.
\newblock In {\em PLDI '91: ACM SIGPLAN 1991 conference on Programming language
  design and implementation}, pages 30--44, New York, NY, USA, 1991. ACM Press.

\bibitem{wu-lcpc08}
Peng Wu, Arun Kejariwal, and Ca{\u{a}}lin Ca{\c{s}}caval.
\newblock Compiler-driven dependence profiling to guide program
  parallelization.
\newblock In {\em LCPC}, pages 232--248, 2008.

\bibitem{zong-hpca08}
Hongtao Zhong, Mojtaba Mehrara, Steve Lieberman, and Scott Mahlke.
\newblock Uncovering hidden loop level parallelism in sequential applications.
\newblock In {\em HPCA}, pages 290--301, 2008.

\bibitem{Zhong:pact03}
Yutao Zhong, Steven~G. Dropsho, and Chen Ding.
\newblock Miss rate prediction across all program inputs.
\newblock In {\em Proc. {PACT}}, 2003.

\bibitem{Zhong:pldi2004}
Yutao Zhong, Maksim Orlovich, Xipeng Shen, and Chen Ding.
\newblock Array regrouping and structure splitting using whole-program
  reference affinity.
\newblock In {\em Proc. {PLDI}}. ACM, 2004.

\end{thebibliography}

\end{document}